\documentclass[%
 reprint,
amsmath,amssymb,
aps,
pra,
]{revtex4-2}

\usepackage{graphicx}
\usepackage{dcolumn}
\usepackage{bm}
\usepackage{tikz}
\usetikzlibrary{quantikz2}
\usepackage{amsmath, amssymb}
\usepackage{listofitems}
\usepackage{xcolor}
\usepackage{placeins}
\usepackage{caption}
\usepackage{float}
\usepackage[hidelinks]{hyperref}

\makeatletter
\let\newfloat\newfloat@ltx
\makeatother
\usepackage{algorithm}
\usepackage{algpseudocode}
\colorlet{myred}{red!80!black}
\colorlet{myblue}{blue!80!black}
\colorlet{mygreen}{green!60!black}
\colorlet{myorange}{orange!70!red!60!black}
\colorlet{mydarkred}{red!30!black}
\colorlet{mydarkblue}{blue!40!black}
\colorlet{mydarkgreen}{green!30!black}

\tikzset{
  >=latex, 
  node/.style={thick,circle,draw=myblue,minimum size=22,inner sep=0.5,outer sep=0.6},
  node in/.style={node,green!20!black,draw=mygreen!30!black,fill=mygreen!25},
  node hidden/.style={node,blue!20!black,draw=myblue!30!black,fill=myblue!20},
  node convol/.style={node,orange!20!black,draw=myorange!30!black,fill=myorange!20},
  node out/.style={node,red!20!black,draw=myred!30!black,fill=myred!20},
  connect/.style={thick,mydarkblue}, 
  connect arrow/.style={-{Latex[length=4,width=3.5]},thick,mydarkblue,shorten <=0.5,shorten >=1},
  node 1/.style={node in}, 
  node 2/.style={node hidden},
  node 3/.style={node out}
}
\def\nstyle{int(\lay<\Nnodlen?min(2,\lay):3)} 

\DeclareMathOperator{\Tr}{Tr}
\DeclareUnicodeCharacter{2212}{-}
\begin{document}

\preprint{APS/123-QED}

\title{Unsupervised Beyond-Standard-Model Event Discovery at the LHC with a Novel Quantum Autoencoder}

\author{Callum Duffy}
\email{callum.duffy.22@ucl.ac.uk}
\author{Mohammad Hassanshahi}%
\author{Marcin Jastrzebski}
\author{Sarah Malik}
\affiliation{%
 University College London, Gower St, London WC1E 6BT\\
}%

\date{\today}

\begin{abstract}

This study explores the potential of unsupervised anomaly detection for identifying physics beyond the Standard Model that may appear at proton collisions at the Large Hadron Collider. We introduce a novel quantum autoencoder circuit ansatz that is specifically designed for this task and demonstrates superior performance compared to previous approaches. To assess its robustness, we evaluate the quantum autoencoder on various types of new physics `signal' events and varying problem sizes. Additionally, we develop classical autoencoders that outperform previously proposed quantum autoencoders but remain outpaced by the new quantum ansatz, despite its significantly reduced number of trainable parameters. Finally, we investigate the properties of quantum autoencoder circuits, focusing on entanglement and magic. We introduce a novel metric in the context of parameterised quantum circuits, stabilizer 2-R\'enyi entropy to quantify magic, along with the previously studied Meyer-Wallach measure for entanglement. Intriguingly, both metrics decreased throughout the training process along with the decrease in the loss function. This appears to suggest that models preferentially learn parameters that reduce these metrics. This study highlights the potential utility of quantum autoencoders in searching for physics beyond the Standard Model at the Large Hadron Collider and opens exciting avenues for further research into the role of entanglement and magic in quantum machine learning more generally.

\end{abstract}

\maketitle

\section{\label{sec:level1}Introduction}

A central objective of high-energy physics (HEP) research at the Large Hadron Collider (LHC) is the identification of potential new physics phenomena in the vast datasets generated by collision events. Unsupervised machine learning techniques, trained on Standard Model (SM) processes, have been utilized to identify whether collision events can be described within the SM framework \cite{Fraser_2022, 10.21468/SciPostPhys.12.1.045, Andreassen_2019, Gonski_2022}. This focus on unbiased machine learning methods allows us to probe a plethora of new physics scenarios \cite{PhysRevD.99.015014, Blance_2019, PhysRevD.105.055006, Govorkova_2022, PhysRevD.101.076015}. However, with the LHC set to transition into its high-luminosity phase, demand for efficient algorithms capable of processing vast volumes of data is paramount. A potential solution for providing efficient computation may come from the field of quantum computation.\newline 

There is compelling motivation for investigating the potential of quantum computing to aid the high-energy physics (HEP) community, with many possible domains of study, such as quantum simulation \cite{PhysRevLett.126.062001, PRXQuantum.4.027001} and the analysis of experimental results \cite{delgado2022quantum}. In this study we apply quantum machine learning (QML) techniques to the task of identifying anomalous collision events at the LHC, which may signal the presence of physics beyond the Standard Model.\newline

QML is the integration of machine learning into the framework of quantum mechanics. Considerable promise has emerged from this field. When compared to its classical counterpart, QML can demonstrate comparable, and in some cases superior, performance in specific tasks \cite{doi:10.1126/science.abk3333}. In instances where a performance gap is observed or proven, the term `quantum advantage' has been introduced. Theoretical demonstrations of quantum advantage, particularly from the perspective of computational complexity \cite{lloyd2013quantum, Liu_2021}, albeit sometimes contrived, have been observed. Furthermore, investigations into sample complexity in generalization from limited training instances \cite{Caro_2022}, along with the exploitation of non-classical correlations within quantum models - a challenge for classical models \cite{Huang_2022, PhysRevX.12.021037}, underscores the anticipated performance advantages of quantum machines.\newline 

Despite QML's promise, there is still much in the field to be resolved, and arguably, one of the largest challenges is trainability. Many of the proposed quantum models suffer from what is known as a barren plateau, where the loss landscape exponentially flattens with model or problem size due to various circuit properties \cite{McClean_2018, Holmes_2022, Cerezo_2021, Thanasilp_2023, Wang_2021}. This, in turn, hinders many gradient-based optimization techniques. Another bottleneck within QML arises due to the infancy of current hardware and the limited number of qubits we can simulate classically.\newline  

QML has already been used for various applications to problems in HEP. Supervised learning approaches have been used for event reconstruction \cite{T_ys_z_2020, T_ys_z_2021, PhysRevD.105.076012, PhysRevD.106.036021, duckett2022reconstructing} and classification tasks \cite{Guan_2021, Wu_2021, Terashi_2021, Blance_2021, Belis_2021} across various problem domains. Similarly, the application of unsupervised learning techniques for detecting beyond-standard model (BSM) signatures has garnered attention \cite{Alvi_2023, woźniak2023quantum, Ngairangbam_2022}.\newline 

Building on these research efforts to detect BSM physics, our work employs a quantum autoencoder (QAE) model. It focuses on identifying a more effective ansatz and exploring larger problem sizes, compared to existing literature, and characterising the inherent properties of these quantum circuits. Previous studies on the QAE successfully identified resonant heavy Higgs signals amidst a QCD $t\bar{t}$ background, leveraging four kinematic features. To expand the scope of QAEs in particle physics, we develop a novel ansatz utilising up to $16$ features and also extend the analysis beyond this signal to include gravitons and scalar bosons.\newline

\section{\label{sec:level3}Autoencoders}
The autoencoder was originally introduced as a neural network designed to reconstruct its input from a compressed latent representation \cite{Rumelhart1986LearningIR}. It compresses data into the lower-dimensional latent representation via a mapping known as the encoder. This compressed representation should capture the input's essential features, enabling a decoder to regenerate an output that faithfully represents the original input. Crucially, the dimensionality of these latent representations is typically less than that of the input data. This reduction is necessary because it allows the model to distil and encode the underlying structures and patterns within complex, high-dimensional datasets. By focusing on these key attributes of the data, the autoencoder can better generalise from the training data to unseen scenarios during testing \cite{fefferman2013testing}. The latent representation produced via the encoder is sometimes called the information bottleneck. We can formally define the learning problem of the autoencoder by the pair of functions $E$ and $D$, which satisfy
\begin{equation}
    argmin_{E,D} \mathbb{E}[\Delta(x,D \circ E(x))],
    \label{eq:autoencoder}
\end{equation}

here $E$, $D$ denote the encoder and decoder respectively, $x$ is the input data and $\Delta$ is the chosen reconstruction loss function.\newline  

Many applications exist for autoencoders, including generative modelling and dimensionality reduction, but our primary interest here is anomaly detection \cite{bank2021autoencoders}. The general principle of using autoencoders for anomaly detection relies on training the model to accurately reconstruct data originating from some non-anomalous distribution. During testing, the model is exposed to both the distribution seen during training and that from an anomalous new distribution. If the autoencoder was trained successfully, input data from the training distribution should be well reconstructed, while data from the anomalous distribution should be poorly reconstructed. From this discrepancy in reconstruction loss, anomalies should be identifiable.\newline 

\subsection{\label{sec:lavel3a} Classical}

\begin{figure}
\begin{tikzpicture}[x=1.2cm,y=1.2cm]
  \large
  \message{^^JNeural network without arrows}
  \readlist\Nnod{5,4,3,2,3,4,5} 
  
  \node[above,align=center,myorange!60!black] at (3,2.4) {$E(\theta, x)$};
  \node[above,align=center,myblue!60!black] at (5,2.4) {$D(\phi, z)$};
  \node[above,align=center,mydarkblue!60!black] at (4,2.0) {$z$};
  \draw[myorange!40,fill=myorange,fill opacity=0.02,rounded corners=2]
    (1.6,-2.7) --++ (0,5.4) --++ (2.8,-1.2) --++ (0,-3) -- cycle;
  \draw[myblue!40,fill=myblue,fill opacity=0.02,rounded corners=2]
    (6.4,-2.7) --++ (0,5.4) --++ (-2.8,-1.2) --++ (0,-3) -- cycle;
  
  \message{^^J  Layer}
  \foreachitem \N \in \Nnod{ 
    \def\lay{\Ncnt} 
    \pgfmathsetmacro\prev{int(\Ncnt-1)} 
    \message{\lay,}
    \foreach \i [evaluate={\y=\N/2-\i+0.5; \x=\lay; \n=\nstyle;}] in {1,...,\N}{ 
      
      \node[node \n,outer sep=0.6] (N\lay-\i) at (\x,\y) {};
      
      \ifnum\lay>1 
        \foreach \j in {1,...,\Nnod[\prev]}{ 
          \draw[connect,white,line width=1.2] (N\prev-\j) -- (N\lay-\i);
          \draw[connect] (N\prev-\j) -- (N\lay-\i);
        }
      \fi 
      
    }
  }
  
  \node[above=0.5,align=center,mygreen!60!black] at (N1-1.90) {input: $x$};
  \node[above=0.5,align=center,myred!60!black] at (N\Nnodlen-1.90) {output: $\hat{x}$};
  
\end{tikzpicture}
\caption{An autoencoder neural network architecture consisting of five input features and a latent space of two. The encoder and decoder each consist of three hidden layers comprising of [4, 3, 2] neurons.}
\label{autoencoder}
\end{figure}
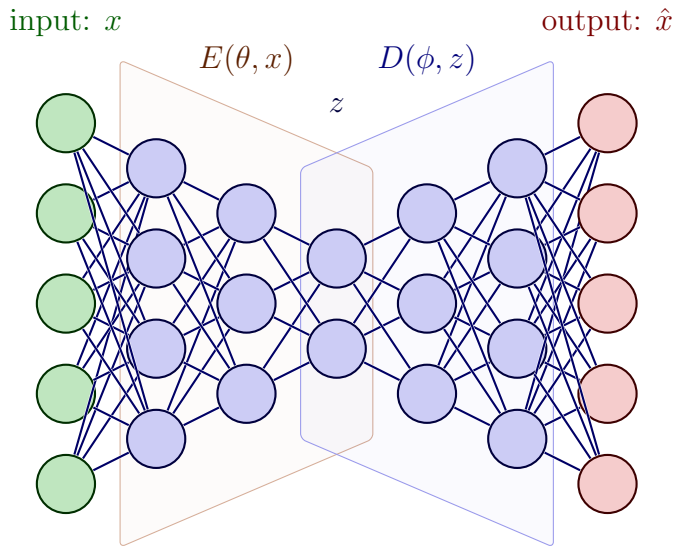

Classical autoencoders (CAEs) are a specific type of neural network architecture with two main components: the encoder and the decoder. The architecture of an autoencoder can often be recognized by the decreasing number of neurons in each subsequent hidden layer of the encoder, followed by a mirror image of these layers in the decoder. The encoder, $E(\theta, x)$, receives inputs $x$, which for our purposes are $n$-dimensional vectors. These are then mapped to a compressed latent representation $z$. The resultant latent representation $z$ is a vector of dimension $k$, which is strictly less than $n$. The decoder, $D(\phi, z)$, then transforms the latent representation $z$ to a vector of dimension $n$ in an attempt to reconstruct the input; we denote this output as $\hat{x}$. This is pictorially represented in Figure \ref{autoencoder}.\newline

An appropriate loss function must be chosen to train an autoencoder, representing an encoding of the problem solution. For this study, we chose the root mean square error (RMSE)
\begin{equation}
    L(x,\hat{x}) = \sqrt{\dfrac{\sum_{i=1}^{i=N}(\hat{x}_i - x_i)^2}{N}}
\end{equation}

where $x_i$ is an input data point, $\hat{x}_i$ the corresponding output from the autoencoder, and $N$ is the total number of points being considered by the loss function.\newline 

Two variants of the CAE were implemented as benchmarks against which to compare the QAE. The first variant is the standard dense neural network, which has varying numbers of hidden layers depending on the problem. The second variant employs a sparse neural network to reduce the number of considered parameters. Due to the all-to-all connectivity between layers of a neural network, the number of parameters grows rapidly with every hidden layer added 
\begin{equation}
    N_{params} = \sum_{i=0}^{L-1}l_i l_{i+1} + \sum_{i=1}^{L}l_i, 
\end{equation}
with $L$ being the number of layers and $l_i$ the number of neurons in layer $i$. The first term calculates the number of weights in the network, and the second term counts the number of biases. Given the limitations of current quantum hardware, which cannot support models with as many parameters as the most advanced classical models, we also use sparse networks for a fairer comparison. The sparse neural networks we constructed contained a number of parameters of the same order of magnitude as those in the quantum models. For this purpose, the jaxpruner library was used, which employs magnitude pruning to mask a user-specified percentage of the lowest magnitude weights during training \cite{jaxpruner}.\newline 

\subsection{\label{sec:lavel3b} Quantum}

The quantum algorithm we consider belongs to the family of variational quantum algorithms (VQAs), which leverage classical and quantum computing resources. VQAs employ parameterized quantum circuits (PQCs), which are constructed using arbitrary rotation and entangling gates. The parametrised gates of the circuit are optimised. Typically, the PQCs are divided into three parts: a feature map, an ansatz and a measurement.\newline 

Feature maps provide a means for embedding classical data into quantum circuits. Often, input data is encoded as the angle within arbitrary rotation gates. Ideally, when mapping data into a Hilbert space, the mapping should exploit some inherent structure in the data such that data belonging to different classes are maximally separated. However, in this study, our focus was not on the feature maps but on the ansatz design.\newline  

The ansatz contains the trainable rotation gates, parametrised by angles $\theta$. By exploiting structural aspects of the data, such as symmetries \cite{ragone2023representation}, or considering the form of the loss function and the corresponding roles of qubits within the circuit, we can strategically inform the design of the ansatz. The selection of a specific ansatz influences the inductive bias of the model and, in turn, the range of functions that the model has access to. We denote an ansatz as $U(\theta)$, a unitary that can be further expressed as a series of sequentially applied unitaries

\begin{equation}
     U(\theta)=U_L(\theta_L)U_{L-1}(\theta_{L-1})...U_2(\theta_2)U_1(\theta_1),
\end{equation}
where
\begin{equation}
    U_i(\theta_i)=\prod_k e^{-i\theta_k H_k}W_k.
\end{equation}
$W_k$ is an unparameterised unitary, $H_k$ is a Hermitian operator and $\theta_i$ is the $i^{th}$ element in $\theta$.\newline

Measurements must be performed to extract information from a PQC. The outcome of these measurements is then fed into an appropriate cost function, which we denote as $C(\theta)$. We can further express the cost function using the construction above as 
\begin{equation}
    C(\theta) = \sum_k f_k (Tr[O_k U(\theta) \rho_k U^{\dag}(\theta)]),
\end{equation}

where $O_k$ is the measurement operator, $U(\theta)$ is the PQC, $\rho_k$ is the input to $U(\theta)$, and $f_k$ represents the classical post-processing required. Cost functions represent a hypersurface consisting of various minima. Ultimately, the goal is to traverse this hypersurface using a chosen optimizer to converge at the global minimum.\newline

\begin{figure}
\begin{quantikz}
\lstick{$\mathcal{H}_{B'} \qquad$} \lstick{$|a\rangle$}& \ghost{H} & \gate[wires=1,2,swap]{} &&\\
\lstick{$\mathcal{H}_B \qquad$} \lstick[2]{$\rho_{in}$} & \gate[2]{U(\theta)}  && \gate[2]{U^\dag(\theta)} & \meter{} \rstick[2]{\qquad $\rho_{out}$}\\
\lstick{$\mathcal{H}_A \qquad$} &&&& \meter{}
\end{quantikz}
\caption{A general schematic of the quantum autoencoder. Input states, $\rho_{in}$, are derived from a feature map and then processed by the unitary ansatze $U(\theta)$, $U^\dag(\theta)$. The Hilbert spaces $\mathcal{H}_A$, $\mathcal{H}_B$ and $\mathcal{H}_{B'}$ are the latent space, the trash space, and the reference space, respectively.}
\label{qae}
\end{figure}
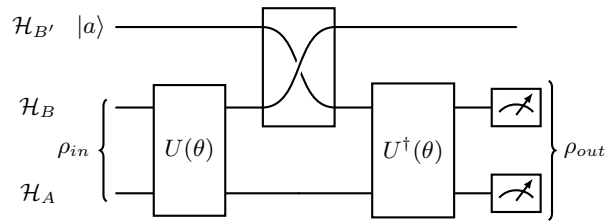

The quantum autoencoder (QAE) is a form of VQA and was conceptualized as the quantum analogue of the CAE \cite{Romero_2017}. Both share a similar feature: the compression of  information. In the case of the QAE, this involves compressing quantum information into a smaller Hilbert space while preserving features of greatest importance. As a learning task, the final reconstructed state $\rho_{out}$ should faithfully represent the initial input state $\rho_{in}$. This accuracy can be checked by calculating the fidelity $F(\rho_{in}, \rho_{out})$.\newline 

A schematic of a QAE can be found in Figure \ref{qae}. The model can be described in terms of three subspaces: $\mathcal{H}_A$ is the latent space, $\mathcal{H}_B$ is sometimes referred to as the `trash' space and represents the space that will be discarded, and $\mathcal{H}_{B'}$ is a reference space required to erase information in $\mathcal{H}_B$. The role of $\mathcal{H}_{B'}$ is described in detail later in this section. The encoder component of this algorithm is described by the circuit $U(\theta)$, which takes trainable parameters as arguments and acts across $\mathcal{H}_A \otimes \mathcal{H}_B$. To achieve data compression, the set of qubits in $\mathcal{H}_B$ is disregarded after the action of $U(\theta)$. One should note here, disregarding $\mathcal{H}_B$ is not done in practise, the reasons for which will be detailed layer. Once we have performed the compression, the next step is to decode the latent representation to the original dimension of the input state, a task performed by a decoder $U^{\dag}(\theta)$. Since $U^{\dag}(\theta)$ acts on the space $\mathcal{H}_A \otimes \mathcal{H}_B$ and the objective of the decoder is to extract useful information solely from the compressed space $\mathcal{H}_A$, the information contained in $\mathcal{H}_B$ must be erased. To achieve this, the information in $\mathcal{H_B}$ is swapped with that in $\mathcal{H_{B'}}$, which contains some reference state $|a\rangle$. This is analogous to erasing the information in $\mathcal{H}_B$. Once this is done and the decoder $U^{\dag}(\theta)$ is applied, we retrieve a state $\rho_{out}$, which, after the parameters $\theta$ have been optimised, should faithfully represent the input state we began with.\newline 

To formalise the above process into a cost function that needs to be minimised, we use
\begin{equation}
    C_1(\theta) = \dfrac{1}{N}\sum_{i=1}^N(1-F(\rho_{i, in}, \rho_{i, out})),
\end{equation}

where $F(\rho_{i, in}, \rho_{i, out})$ is the fidelity between the input $\rho_{i, in}$ and output $\rho_{i, out}$ states, and the input data is encoded into input states $\rho_{i, in}$.\newline 

It can be shown that the quantum autoencoder architecture can be simplified to a form which does not make use of $U^{\dag}(\theta)$ explicitly \cite{Ngairangbam_2022, Romero_2017}. A simple explanation of this method begins with identifying an encoder unitary which performs perfect compression. Specifically, for all input states it produces: 
\begin{equation}
    U(\theta)|\psi_i\rangle_{AB} = |\psi_i^c\rangle_A \otimes |a\rangle_B,
\label{eq:perf_compression}
\end{equation}
where $|\psi_i^c\rangle_A$ is the compressed state, and $|a\rangle_B$ is the reference state. Upon the action of $U^{\dag}(\theta)$, $|\psi_i\rangle_{AB}$ will be recovered for all $i$ since the unitary $U(\theta)$ has been able to produce the reference state $|a\rangle$ in $\mathcal{H}_B$ for all input states. As a result, for the learning task, it suffices to consider whether, after performing $U(\theta)$ the resultant state in $\mathcal{H}_B$ matches that of the reference state in $\mathcal{H}_{B'}$. To achieve this, we compute the fidelity between the two states. From the above, we can re-state the cost function as 
\begin{equation}
    C_2(\theta) = \dfrac{1}{N}\sum_{i=1}^N  (1 - F(\Tr_A[U(\theta)\rho_{i,in}U^{\dag}(\theta)], |a\rangle\langle a|_{B'})).
\label{eq:loss_fn}
\end{equation}

\begin{figure}
\begin{quantikz}
\lstick[2]{$\mathcal{H}_A$}& \gate[4]{F(x)} & \gate[4]{U(\theta)} &&&& \\
& &&&&&\\
\lstick[2]{$\mathcal{H}_B$}&&& \swap{2} &&&\\
&&&& \swap{2} &&\\
\lstick[2]{$\mathcal{H}_{B'}$}&&& \targX{} &&&\\
&&&& \targX{} && \\
&& \gate{H} & \ctrl{-2} & \ctrl{-1} & \gate{H} & \meter{} 
\end{quantikz}
\caption{A diagram showing the form of the autoencoder used in practise. Here, we have four input qubits and a latent space of two qubits. Initial input states $|\psi_i\rangle$ are formed from a feature map $F(x)$ embedding classical data in a Hilbert space. States created by $F(x)$ are then passed into the chosen ansatz $U(\theta)$. Finally, a SWAP test is performed between the spaces $\mathcal{H}_B$ and $\mathcal{H}_{B'}$ to extract information from the circuit.}
\label{circ:swap_test}
\end{figure}
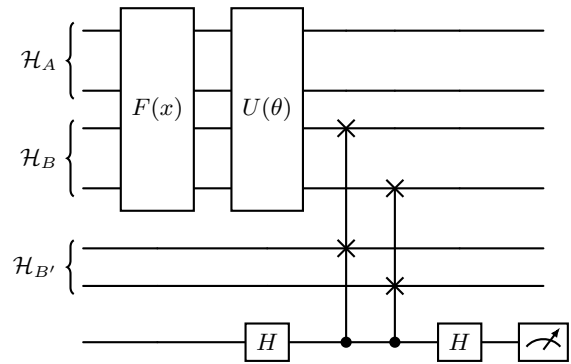

For implementing the fidelity measurement in practice between $\mathcal{H}_B$ and $\mathcal{H}_{B'}$, the swap test is used and is shown in Figure \ref{circ:swap_test} after the action of a feature map $F(x)$ and ansatz $U(\theta)$. The reference state is chosen to be the all-zero state.\newline 

For this paper, we propose an ansatz well suited to the task of a QAE. To demonstrate its effectiveness, we compare it to two popular ansatz. In all parameterised circuits used, the rotation gates are kept the same, but the entangling structure differs.\newline

\begin{figure}
\begin{quantikz}
& \gate{R_y(\theta)} & \ctrl{1} & \ctrl{2} & \ctrl{3} &&& \\
& \gate{R_y(\theta)} & \targ{} &&& \ctrl{1} & \ctrl{2} & \\
& \gate{R_y(\theta)} && \targ{} && \targ{} && \ctrl{1}\\
& \gate{R_y(\theta)} &&& \targ{} && \targ{} & \targ{}\\
\end{quantikz}
\caption{The all-to-all entangling ansatz, which contains parameterised Pauli-Y rotation gates followed by CNOT gates that entangle all pairs of qubits to one another.}
\label{circ:all-to-all}
\end{figure}
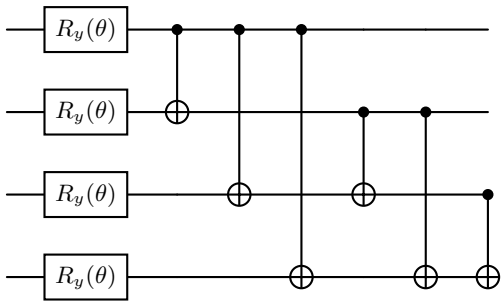

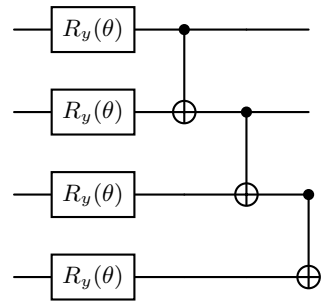
\begin{figure}
\begin{quantikz}
& \gate{R_y(\theta)} & \ctrl{1} &&\\
& \gate{R_y(\theta)} & \targ{} & \ctrl{1} &\\
& \gate{R_y(\theta)} && \targ{} & \ctrl{1}\\
& \gate{R_y(\theta)} &&& \targ{} \\
\end{quantikz}
\caption{The HEA, which contains parameterised Pauli-Y rotation gates followed by CNOT gates that entangle neighbouring pairs of qubits to respect the qubit connectivity of real quantum devices.}
\label{circ:hae}
\end{figure}

The first of these ansatz can be seen in Figure \ref{circ:all-to-all}, depicting all-to-all entangling CNOT gates in the case of four qubits, to spread correlations amongst all qubits. The number of CNOT gates required scales quadratically with the number of qubits. This circuit was considered in previous work when dealing with anomaly detection in particle physics \cite{Ngairangbam_2022}, we will refer to this ansatz from here on as the original ansatz. The second ansatz, the hardware efficient ansatz (HEA), contains CNOT gates between nearest neighbour qubits. A four-qubit example of this ansatz can be seen in Figure \ref{circ:hae}. The idea behind this structure is to respect the connectivity of many near-term quantum devices to minimise the effect of hardware noise \cite{leone2022practical}.\newline  

\begin{figure}
\begin{quantikz}
&\targ{}\gategroup[wires=4,steps
=3,style={dashed,
rounded corners,inner sep=6pt}]{repeat $\times \ell$}& \gate{R_y(\theta)} && \ctrl{3} &&  \rstick[2]{latent}\\
&& \targ{} & \gate{R_y(\theta)} && \ctrl{1} & \\
&& \ctrl{-1} & \gate{R_y(\theta)} && \targ{} &  \rstick[2]{trash}\\
& \ctrl{-3} & \gate{R_y(\theta)} && \targ{} && 
\end{quantikz}
\caption{The newly proposed ansatz. Initially CNOT gates link every qubit in the trash space to those in the latent space, follow by parameterised arbitrary Pauli-Y rotation gates. From here a second set of CNOT gates which act on every qubit in the latent space with target bits lying in the trash space. This ansatz is aware of which qubits belong to $\mathcal{H}_A$ and which to $\mathcal{H}_B$, adapting where CNOT gates act accordingly.}
\label{circ:new}
\end{figure}
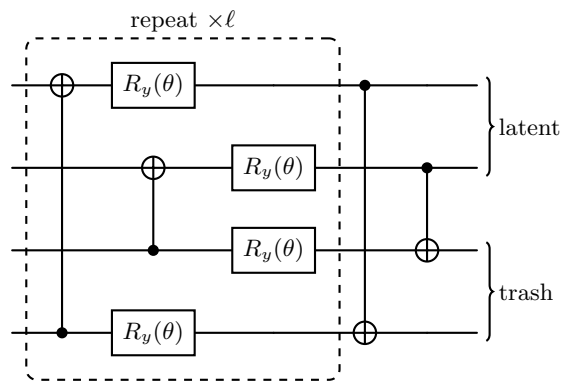

The ansatz we propose contains CNOT gates both before and after the rotation gates, as shown in Figure \ref{circ:new}, for the specific case of four qubits, with a latent and trash space of two qubits. Recent literature has stressed the importance of inductive biases in quantum machine learning models, whereby restricting the class of functions the model has access to can improve generalisation performance \cite{ragone2023representation}. The choice of a unitary ansatz directly influences the class of functions the model can produce. Here, the structure of the circuit is informed by the role of each qubit, such as whether it belongs to the latent space or if its information will be discarded. The initial set of CNOT gates aims to transfer information contained in qubits that belong to $\mathcal{H}_B$ to qubits that will remain in the latent space $\mathcal{H}_A$. Here, we have chosen the control bit to lie in $\mathcal{H}_B$ and the target bit in $\mathcal{H}_A$. This choice only becomes relevant for the special case of the input state being perfectly compressed into $\mathcal{H}_A$, leaving $\mathcal{H}_B$ in the all zero-state. Subsequent CNOT gates will then no longer alter the state.\newline

Both sets of entangling gates (before and after rotations) have been found to be crucial for the enhanced performance of the ansatz. If one aims to have $\ell$ layers of $R_y(\theta)$ gates, the CNOT gates performing information transfer are repeated $\ell$ times, with the final layer also containing CNOT gates which erase information in the qubits to be discarded. This is a step towards constructing a more standardised ansatz for QAEs, which has been highlighted as an important goal in the literature \cite{Caro_2022}. We focus on producing an ansatz with fewer CNOT gates compared to the all-to-all ansatz and select those considered most useful for the computation at hand. Prioritising the transfer of information between the latent and trash space to bias the model to compress the most valuable information into the latent space. Further details on how this ansatz should be implemented for a chosen latent space size, number of layers and qubits are in Appendix \ref{appx:pseudocode}.\newline

\section{\label{sec:level2}Circuit Properties}
In addition to analysing the anomaly detection performance of the circuits discussed, two information-theoretic properties of interest are investigated. These are: entanglement and magic. Both entanglement and magic provide the means to quantify the amount of quantum resources we are utilising, which in turn, helps gauge whether the circuits being used are beyond the capabilities of classical computing.
While it is currently not possible to conclusively determine that a specific quantum circuit cannot be simulated classically simply by calculating these metrics, these metrics can still provide useful insights. More concretely, both of these resources need to be used but, obtaining maximal entanglement or magic available leads to states which are not conducive to obtaining quantum speedups \cite{PRXQuantum.3.020333,PhysRevLett.102.190501}.\newline 

\subsection{\label{subsec:level2b}Entanglement}
Generating entanglement is a necessary but not sufficient condition for quantum circuits that aim to achieve a quantum advantage. In quantum machine learning, entanglement is often employed to capture non-trivial correlations in data \cite{Schuld_2020, Kandala_2017}. 
CNOT and CZ gates are commonly used to realise entanglement in quantum circuits.\newline

To empirically measure entanglement, the Meyer-Wallach entanglement measure \cite{Meyer_2002} can be used. This is a global measure of multi-particle entanglement for pure states, aggregating bipartite entanglement measures across different subsystems. Often denoted as Q, the Meyer-Wallach entanglement measure can be defined as follows: Consider an $n$ qubit system with a linear map $l_j(b)$ acting on the computational basis states 
\begin{equation}
    l_j(b)|b_1 ... b_n\rangle = \delta_{bb_j}|b_1...\hat{b}_j ... b_n \rangle,
\end{equation}

where $b_j \in \{0,1\}$ and $\hat{}$ indicates qubit j being absent. The entanglement measure can then be defined as 
\begin{equation}
Q(|\psi\rangle) = \dfrac{4}{n}\sum_{j=1}^{n}D(l_j(0)|\psi\rangle,l_j(1)|\psi\rangle),
\end{equation}
with 
\begin{equation}
    D(|u\rangle,|v\rangle) = \dfrac{1}{2}\sum|u_i v_j - u_jv_i|^2,
\end{equation}

$|u\rangle=\sum u_i |i\rangle$ and $|v\rangle = \sum v_i|i\rangle$. $D$ can be seen as a distance operator. The measure defined here has the following properties:
\begin{itemize}
    \item Invariant under local unitaries.
    \item $0 \leq Q \leq 1$.
    \item $Q(|\psi\rangle) = 0$ iff $|\psi\rangle$ is a product state.
    \item $Q(|\psi\rangle) = 1$ iff $|\psi\rangle$ is a maximally entangled state.
\end{itemize}

Previous work has used this measure to characterise entanglement in error correcting-codes and quantum phase transitions \cite{Meyer_2002, Somma_2004}. This definition has been extended to characterise parametrised quantum circuits by taking the average of the Meyer-Wallach entanglement measure over a collection of sampled parameters, more formally \cite{Sim_2019}
\begin{equation}
    \bar{Q}(|\psi(\theta)\rangle) = \dfrac{1}{|S|}\sum_{\theta \in S}Q(|\psi(\theta)\rangle),
\end{equation}
where $S\in\{\theta\}$ is the set of sampled circuit parameters. Here only the ansatz of the quantum circuit is considered.\newline

For this study, we consider the circuit as a whole, including both the ansatz and feature map. Focusing solely on the ansatz neglects that it acts on product or entangled states generated via the feature map. As a result, we study the distribution of global entanglement as a function of both parameters $\theta$ and data $x$, $Q(|\psi(\theta, x)\rangle)$. From here, there are two distributions of interest to us, the first where we fix $\theta$ to be the set of parameters found after training, giving rise to a distribution over $Q(|\psi(x)_{\theta_{trained}}\rangle)$. This will give us insight into the global entanglement distribution for an optimised circuit over a set of sampled data points. We will also define the mean of this distribution as
\begin{equation}
    \bar{Q}(|\psi(x)|_{\theta_{trained}})\rangle) = \dfrac{1}{|D|}\sum_{x \in D}Q(|\psi(x)|_{\theta_{trained}}\rangle),
\label{eq:entanglement_data_mean}
\end{equation}
where $D\in \{x\}$ is the set of sampled data points.\newline 

The second distribution we will define is that over both $x$ and $\theta$, $Q(|\psi(x,\theta)\rangle)$. 
We also define the mean global entanglement of this distribution by
\begin{equation}
    \bar{Q}(|\psi(x, \theta)\rangle) = \dfrac{1}{|S||D|}\sum_{\theta \in S}\sum_{x\in D}Q(|\psi(\theta, x)\rangle),
\label{eq:entanglement_data_param_mean}
\end{equation}
where $\theta$ is sampled from the uniform distribution. By comparing equations \ref{eq:entanglement_data_mean} and \ref{eq:entanglement_data_param_mean} we will be able to ascertain whether circuits with optimized parameters preferentially select states with high or low global entanglement.\newline

\subsection{\label{subsec:level2c}Magic}

Like entanglement, the magic a quantum circuit possesses is a necessary but not sufficient condition to avoid classical simulatability. The notion of magic arises from the fact that by the Gottesman-Knill theorem, circuits consisting of gates from the Clifford group can be simulated efficiently classically even if they are created with high levels of entanglement \cite{gottesman1998heisenberg}. Therefore, creating circuits that are not efficiently simulatable requires the using non-Clifford gates. These non-Clifford gates can come in the form of $T$ and Toffoli gates. Implementing these resources gives quantum computers the ability to outperform classical machines, which has been dubbed as providing the circuit with `magic'. Subsequently, a resource theory for magic has been formulated \cite{PhysRevLett.128.050402, PhysRevA.83.032317}. We can define a measure of magic via the stabilizer $2$-R\'enyi entropy
\begin{equation}
    M_2 (|\psi\rangle) := - \log_2 W(|\psi\rangle) - S_2(|\psi\rangle) - \log_2(d),
    \label{eq:magic}
\end{equation}
where $W(|\psi\rangle) := \Tr(Q|\psi\rangle^{\otimes 4})$, $Q := d^{-2}\sum_P P^{\otimes 4}$ and $d=2^n$, the sum is taken over all multi-qubit strings of the Pauli operators to four copies of the state and $S_2(|\psi\rangle) = -\log_2(\Tr(|\psi\rangle^2))$ is the $2$-R\'enyi entropy. This measure was originally proposed in ref. \cite{PhysRevLett.128.050402}, along with an experimental protocol; later, this protocol was improved by ref. \cite{Oliviero_2022} to only include randomized single-qubit measurements rather than global multi-qubit measurements. For this work, we calculated 
the analytic value of magic via equation \ref{eq:magic} with statevector simulation and did not use the experimental protocols.\newline 

This measure of magic has the following properties:
\begin{itemize}
    \item $M_2(|\psi\rangle)=0$, in the case all the gates in the circuit come from the Clifford group.
    \item $M_2(C|\psi\rangle)=M_2(|\psi\rangle)$, invariant under Clifford rotations.
    \item $0<M_2(|\psi\rangle) \leq \log(d+1) -\log(2)$.
\end{itemize}

In the same spirit as global entanglement measures, we propose to take averages over a set of parameters and data points to gain intuition about the magic generated by our quantum circuits. We construct similar expressions for the magic measure
\begin{equation}
    \bar{M}_2(|\psi(x)|_{\theta_{trained}}\rangle) = \dfrac{1}{|D|}\sum_{x\in D}M_2(|\psi(x)|_{\theta_{trained}}\rangle).
    \label{eq:magic_data_mean}
\end{equation}
and
\begin{equation}
    \bar{M}_2(|\psi(x,\theta)\rangle) = \dfrac{1}{|S||D|}\sum_{\theta \in S}\sum_{x\in D}M_2(|\psi(\theta, x)\rangle).
    \label{eq:magic_data_param_mean}
\end{equation}

\section{\label{sec:level4}Datasets}

The datasets under consideration consist of kinematic information describing proton-proton particle collisions. The two main processes being examined are events well characterised by QCD and those from beyond standard model (BSM) processes. Events originating from BSM processes cannot be well described within the theory of QCD and, for this study, are the signals of interest. Meanwhile, events described by QCD theory will be designated as the in-distribution background data.\newline 

The form of the data we are considering is kinematic data from particle collisions, thus classical. However, it is important to realise the classical data used here originates from quantum processes.\newline 

By considering two separate datasets and feature maps, we hope to demonstrate the performance advantages of the newly proposed ansatz design, which could be broadly applicable, at least in the context of HEP data. We will now introduce the two datasets we studied, which consist of four signals in total.\newline 

\subsection{\label{subsec:level4a} Heavy Higgs}
 
This dataset is taken from \cite{Ngairangbam_2022}, which considered the background QCD process of $t\bar{t}$ production $pp\rightarrow t\bar{t}$, with signals originating from scalar resonance production $pp \rightarrow H \rightarrow t\bar{t}$. The signal of interest here involves a heavy Higgs of mass $m_H = 2.5$Tev. All top quarks decay to a bottom quark and a W boson, which decays to a muon. Both events were generated according to a centre of mass energy of $14$Tev via MadGraph5 aMC@NLO \cite{Alwall_2014}. Showering and hadronization were taken care of by Pythia8 \cite{Sj_strand_2015}, while Delphes3 was utilized for detector simulation \cite{de_Favereau_2014}.\newline 

Object reconstruction was implemented with the anti-$k_t$ algorithm with a jet radius R=$0.5$. Bottom jets originating from signal events were required to have $p_T^b > 30$Gev with isolation criteria of $R=0.5$. The variables extracted from this process were:
\begin{itemize}
    \item Two lepton transverse momenta $p_T^{l_1}$, $p_T^{l_2}$.
    \item Angle between two leptons $\theta_{l}$.
    \item Two quark transverse momenta $p_T^{q_1}$, $p_T^{q_2}$.
    \item Angle between two quarks $\theta_{q}$.
    \item Transverse energy $E_T$.
    \item Two lepton angular separations $dR_{l_1}$, $dR_{l_2}$.
\end{itemize}

\begin{figure}
\begin{quantikz}
& \gate{R_x(x_1)} & \\
& \gate{R_x(x_2)} & \\
&\phantom{C}\vdots\phantom{C}&\\
& \gate{R_x(x_1)} & \\
& \gate{R_x(x_2)} & \\
\end{quantikz}
\caption{A feature map for the Heavy Higgs dataset consisting solely of $R_x$ rotation gates to embed features $x$.}
\label{circ:higgs_feat_map}
\end{figure}
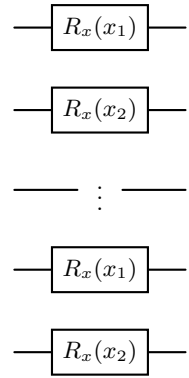

Each variable was mapped to the interval $[0, \pi]$ for QAEs and $[0, 1]$ for CAEs, where non-angular variables are fixed between $[0,1000]$ by adding two artificial extreme data points before scaling. The faux data is then removed. A reasonably simple feature map was implemented to embed the data from this dataset, consisting solely of $R_x(\theta)$ rotation gates seen in Figure \ref{circ:higgs_feat_map}.\newline 

The previous study, which utilised this dataset as well as the QAE, used a subset consisting of four kinematic variables $\{E_T, p_T^{b_1}, p_T^{l_1}, p_T^{l_2}\}$. For our study, we studied three subsets of increasing size:
\begin{itemize}
    \item $\{E_T, \theta_l, p_T^{l_1}, p_T^{l_2}\}$,
    \item $\{E_T, p_T^{b_1}, p_T^{l_1}, p_T^{l_2}, \theta_l, p_T^{b_2}\}$,
    \item $\{E_T, p_T^{b_1}, p_T^{l_1}, p_T^{l_2}, \theta_l, p_T^{b_2}, \theta_b, dR_{l_1}\}$,
\end{itemize}

with the largest subset containing eight features.\newline

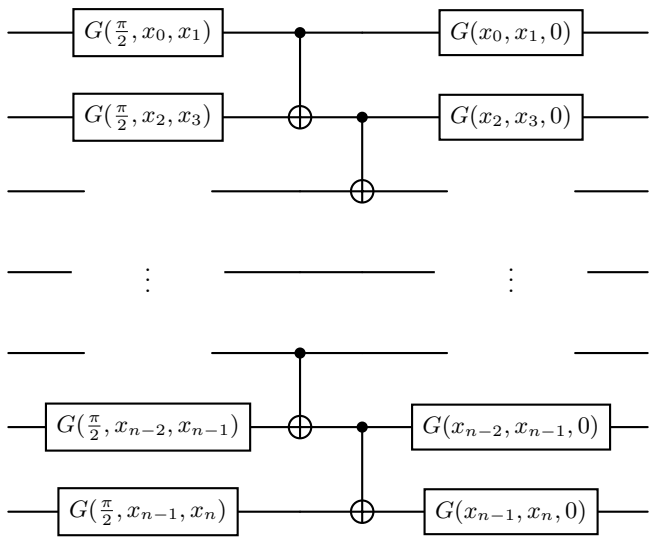
\begin{figure}
\begin{quantikz}
& \gate{G(\frac{\pi}{2}, x_0, x_1)} & \ctrl{1} && \gate{G(x_0, x_1, 0)}&\\
& \gate{G(\frac{\pi}{2}, x_2, x_3)} & \targ{} & \ctrl{1} & \gate{G(x_2, x_3, 0)}&\\
&\phantom{G(\frac{\pi}{2}, x_2, x_3)} && \targ{} &\phantom{G(\frac{\pi}{2}, x_2, x_3)}&\\
&\gate[1,style={fill=white,draw=white,text width=2cm}]{\vdots} &&& \gate[1,style={fill=white,draw=white,text width=2cm}]{\vdots} &\\
&\phantom{G(\frac{\pi}{2}, x_2, x_3)} & \ctrl{1} && \phantom{G(\frac{\pi}{2}, x_2, x_3)} &\\
& \gate{G(\frac{\pi}{2}, x_{n-2}, x_{n-1})} & \targ{} & \ctrl{1} & \gate{G(x_{n-2}, x_{n-1}, 0)}&\\
& \gate{G(\frac{\pi}{2}, x_{n-1}, x_{n})} && \targ{} & \gate{G(x_{n-1}, x_{n}, 0)}&\\
\end{quantikz}
\caption{A feature map for the graviton and scalar boson dataset consisting solely $G(\phi, \theta, \omega)$ rotation gates ($G(\phi, \theta, \omega) = R_z(\omega) R_y(\theta) R_z(\phi)$) and nearest neighbour entangling gates.}
\label{circ:graviton_feat_map}
\end{figure}

\begin{figure*}[t!]
\begin{tabular}{ccc}
     \includegraphics[width=0.32\textwidth]{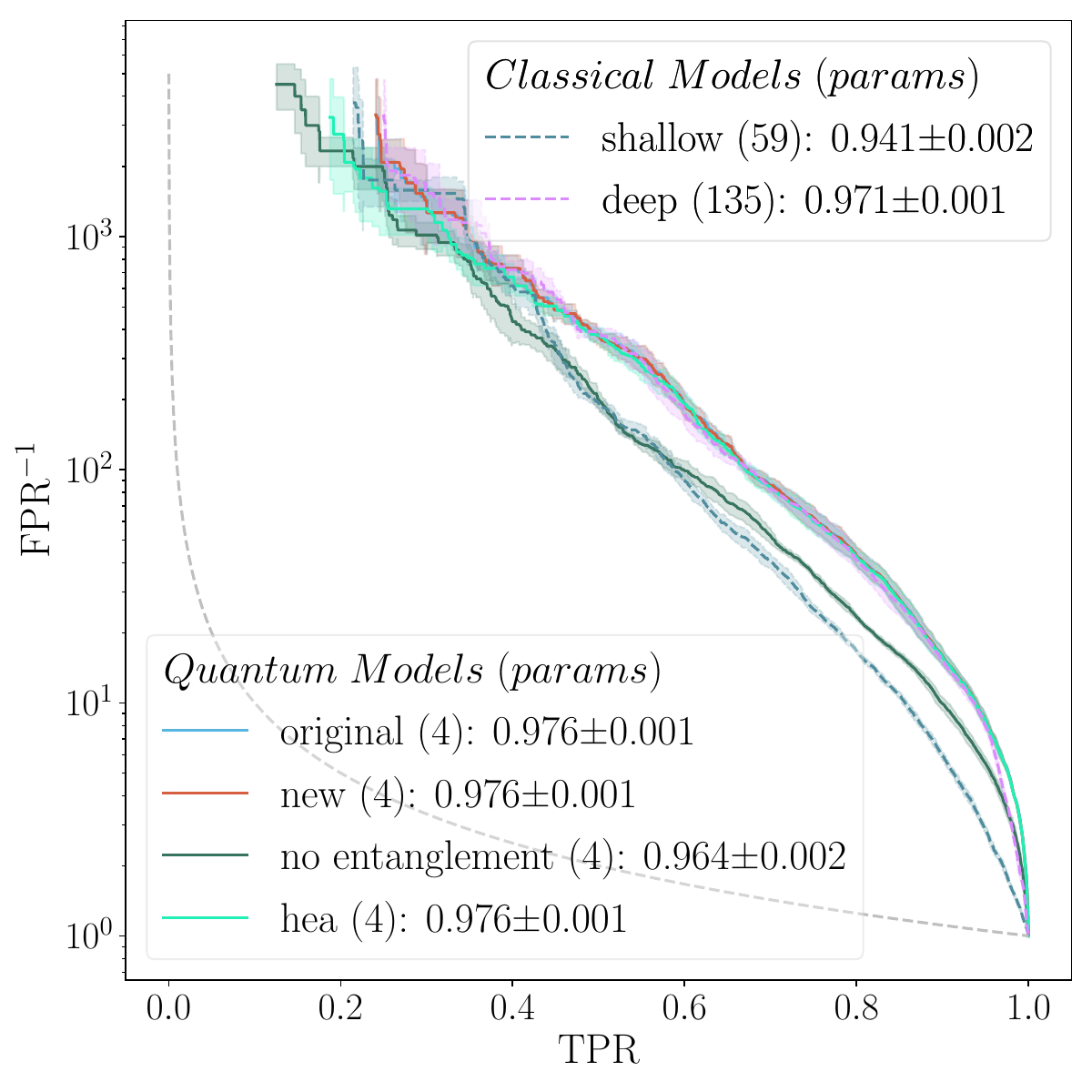}& \includegraphics[width=0.32\textwidth]{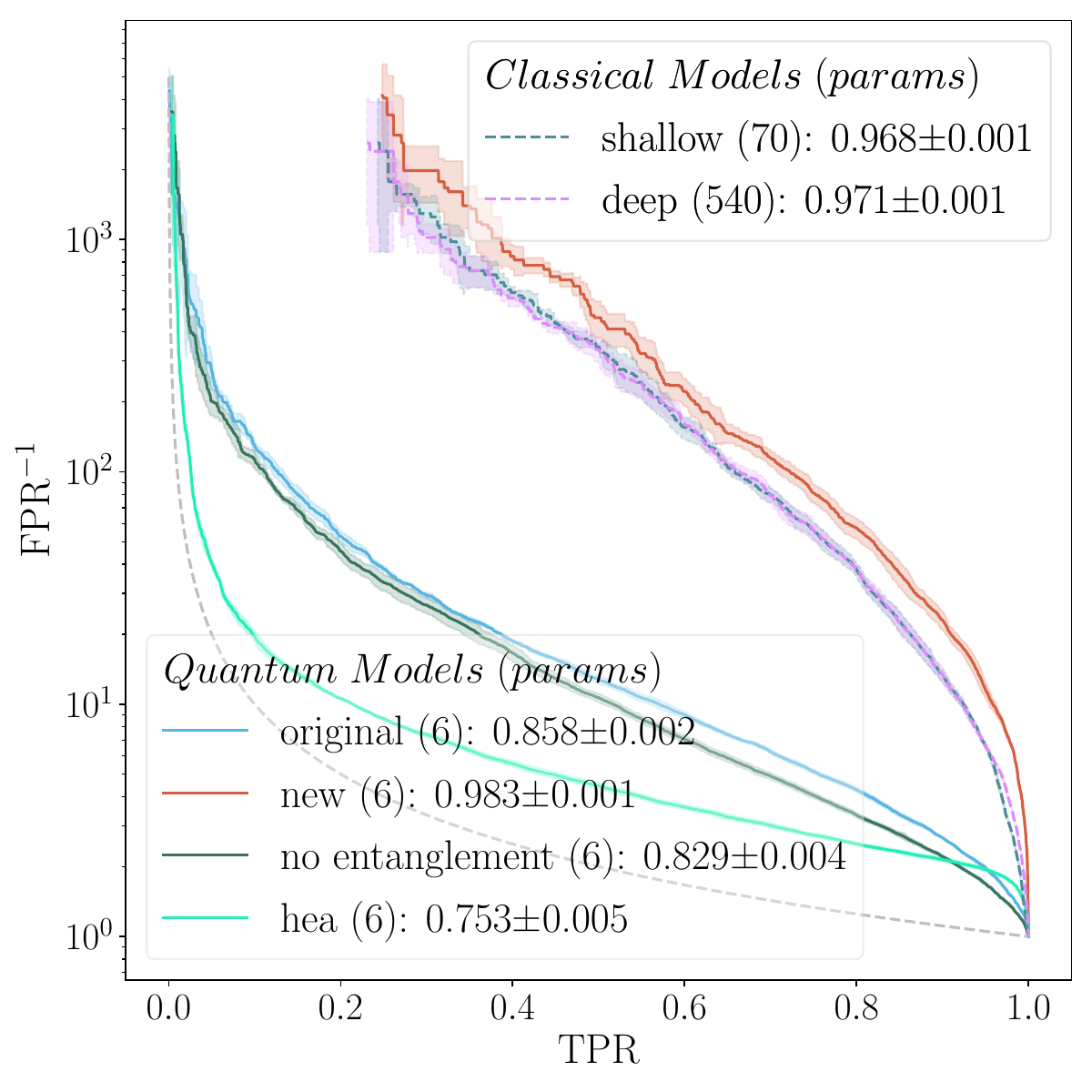} & \includegraphics[width=0.32\textwidth]{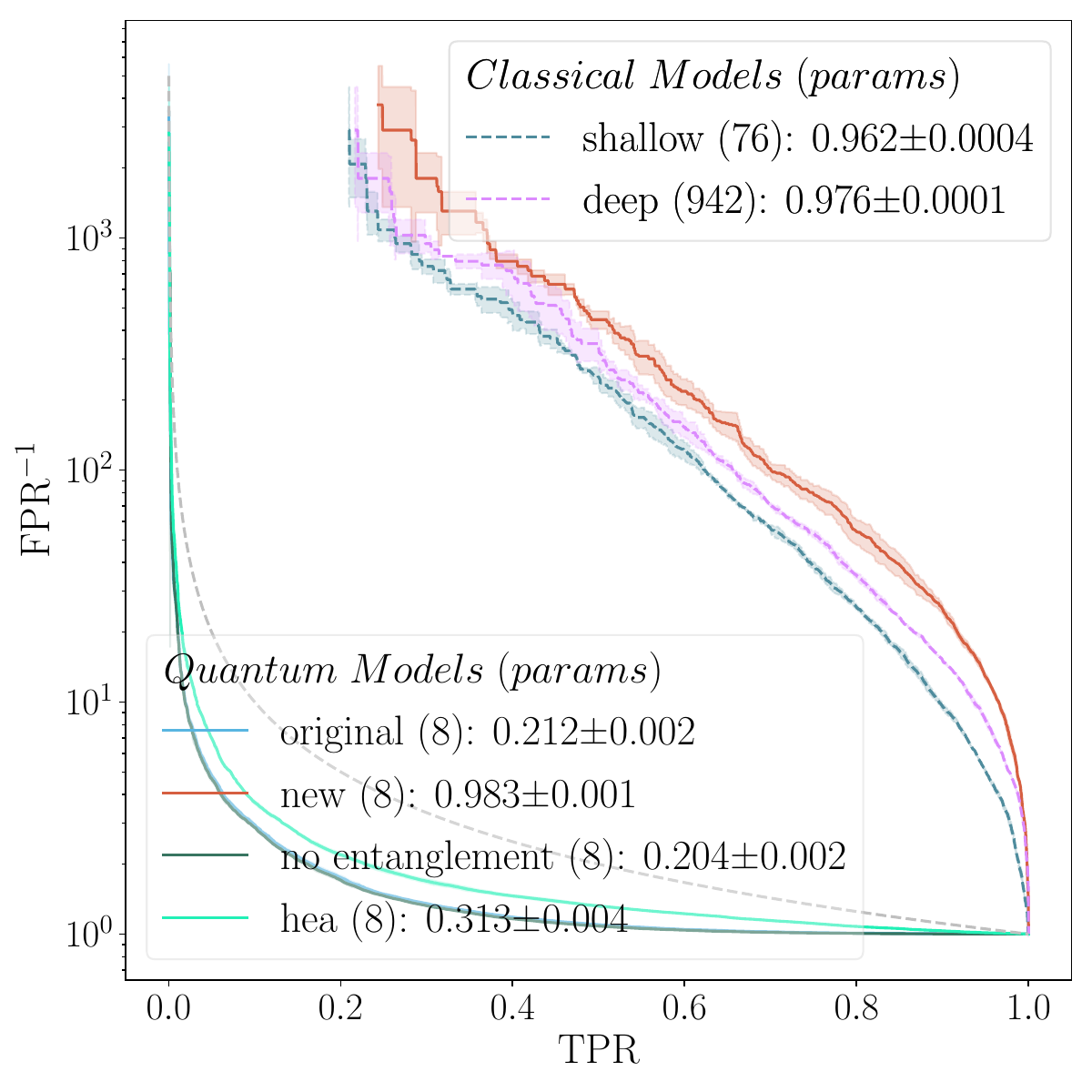}  \\
     (a) $4$ features &  (b) $6$ features & (c) $8$ features \\[6pt]
\end{tabular}
    \caption{Each subplot depicts the ROC curve for a collection of quantum and classical autoencoders on test data for a given set of input features. The grey dashed line represents a model performing random guessing. Panel (a) four input features with kinematic variables $\{E_T, p^{b_1}_{T}, p^{l_1}_{T}, p^{l_1}_{T}\}$ with five folds of test data. Panel (b) six input features with kinematic variables $\{E_T, p^{b_1}_{T}, p^{l_1}_{T}, p^{l_1}_{T}, \theta_l, p^{b_1}_{T}\}$, with $3$ folds of test data. Panel (c) eight input features with kinematic variables $\{E_T, p^{b_1}_{T}, p^{l_1}_{T}, p^{l_1}_{T}, \theta_l, p^{b_1}_{T}, \theta_b, dR_{l1}\}$, with $3$ folds of test data.}
    \label{fig:higgs_roc}
\end{figure*}

\subsection{\label{subsec:level4b} Randall-Sundrum Gravitons and ZZZ Scalar Bosons}

A previous study into anomaly detection with quantum kernels created the second dataset we considered \cite{woźniak2023quantum}. The study was yet again into proton-proton collisions at a centre of mass energy of $13$Tev, but in this case the BSM processes were:
\begin{itemize}
    \item Narrow Randall-Sundrum graviton decaying to two $W$-bosons, Narrow $G \rightarrow WW$ \cite{Randall_1999}. The width of the graviton was set to a negligible value by fixing $\kappa m_G$ to $0.01$, where $\kappa$ is a dimensionless coupling parameter \cite{Bijnens_2001}.
    
    \item Broad Randall-Sundrum graviton decaying to two W-bosons, Broad $G \rightarrow WW$ \cite{Randall_1999}. The parameter $\kappa m_G$ was set to $2.5$, resulting in a width-to-mean ratio of $~35\%$ for $m_{jj}$ after detector effects.
    
    \item Scalar boson $A$ decaying to a Higgs and a $Z$ boson, $A \rightarrow HZ \rightarrow ZZZ$.
\end{itemize}

All W and Z bosons are set to decay to quark pairs, producing all-jet final states. The background processes simulated were QCD multi-jet production.\newline 

The Delphes3 library was used to simulate detector effects mimicking the CMS detector description, and a proton in-time pileup of $40$. The size of the dataset produced equates to an integrated luminosity of $64\text{fb}^{-1}$.\newline 

A particle-flow algorithm followed by the anti-$k_t$ clustering algorithm (R=0.8) was employed to process the events. To make the dataset representative of LHC event selection, only dijet events satisfying the invariant mass condition $m_{jj} > 1260$GeV are selected. Moreover, dijets must satisfy $p_T > 200$GeV and $|\nu|<2.4$. Events comprise the two highest $p_T$ jets, with each jet further comprising the 100 highest $p_T$ particles zero-padded when necessary. The kinematic variables used to describe each particle within a jet are $p_T$, $\eta$ and $\phi$.\newline 

Since the dataset described above results in a single jet represented by a $100\times 3$ matrix, dimensionality reduction is required to allow current NISQ devices to process the data \cite{Preskill2018quantumcomputingin}. The method implemented by Ref. \cite{woźniak2023quantum} was a 1D convolutional autoencoder, to respect the structure of the jet data. Jet objects from the QCD background data in the sideband region are compressed, resulting in a reduced dijet dataset of $2n$ features, with $n$ being the dimension of the latent representation of each jet. They argue that dimensionality reduction of this nature can partially retain non-linear correlations in the latent space compared to the likes of other dimensionality reduction methods such as PCA.\newline 

The feature map used, as seen in Figure \ref{circ:graviton_feat_map}, is the same as in the original study to focus on the effects of the newly proposed ansatz.\newline

\section{\label{sec:level5}Results and Discussion}

The following QAE models presented were created using Pennylane \cite{bergholm2022pennylane}, while the CAEs were built using the Jax ecosystem, utilising Flax and Optax \cite{jax2018github,flax2020github,deepmind2020jax}. For training, $1000$ background samples were used, and for testing, $10,000$ samples were used equally split between signal and background sources.\newline 

The classical models presented have been found from a randomised grid search at each input feature size. Shallow models were found from the best-performing models containing equal to or fewer than $100$ parameters. Deep models could exceed the limit of $100$ parameters but were restricted to fewer than $1000$ parameters so as not to enter the over-parameterised regime. These models were trained for $500$ epochs with the Adam optimizer. We refer the reader to appendix \ref{appx:grid_search} for details on the exact nature of the grid search.\newline 

In the case of the quantum models, each was trained with the Adam optimizer. Unless otherwise stated, a single-layer ansatz was used. A randomised grid search was used to determine the model when more than a single layer ansatz was seen. Further details on the QAEs used can be found in appendix \ref{appx:model_details} along with CAEs.\newline

\subsection{\label{sec:level5a} Performance}

\begin{figure*}[t!]
\begin{tabular}{ccc}
     \includegraphics[width=0.32\textwidth]{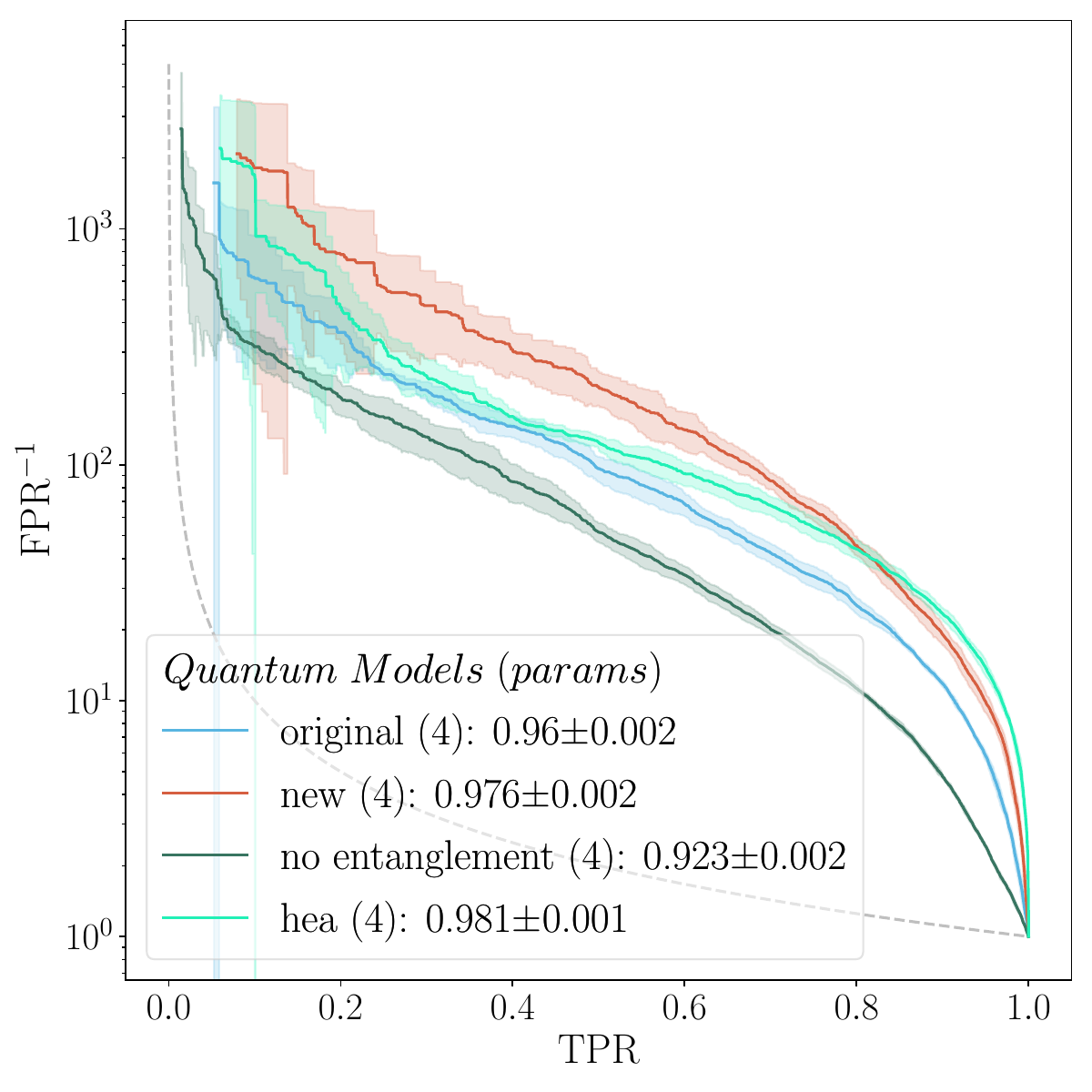}& \includegraphics[width=0.32\textwidth]{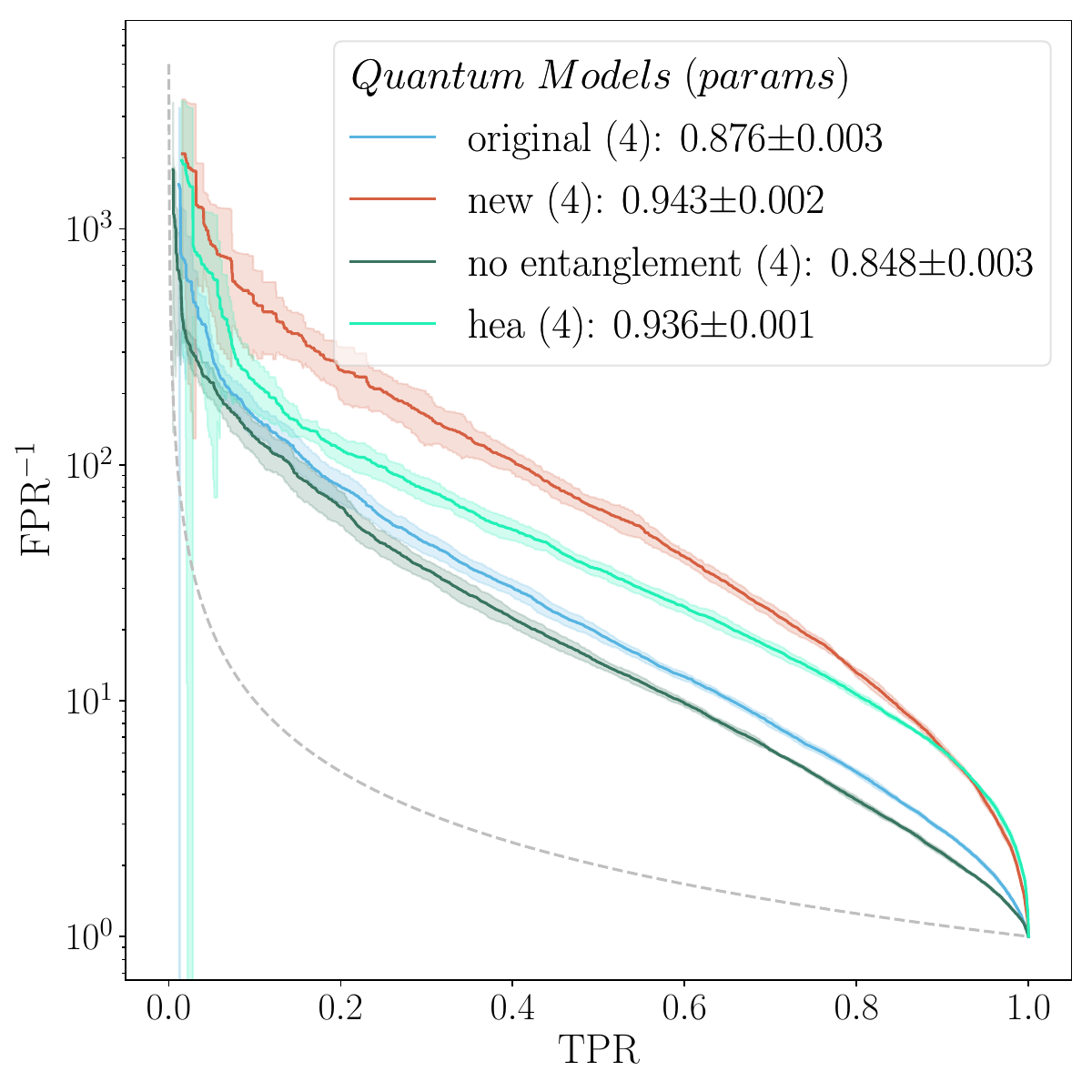} & \includegraphics[width=0.32\textwidth]{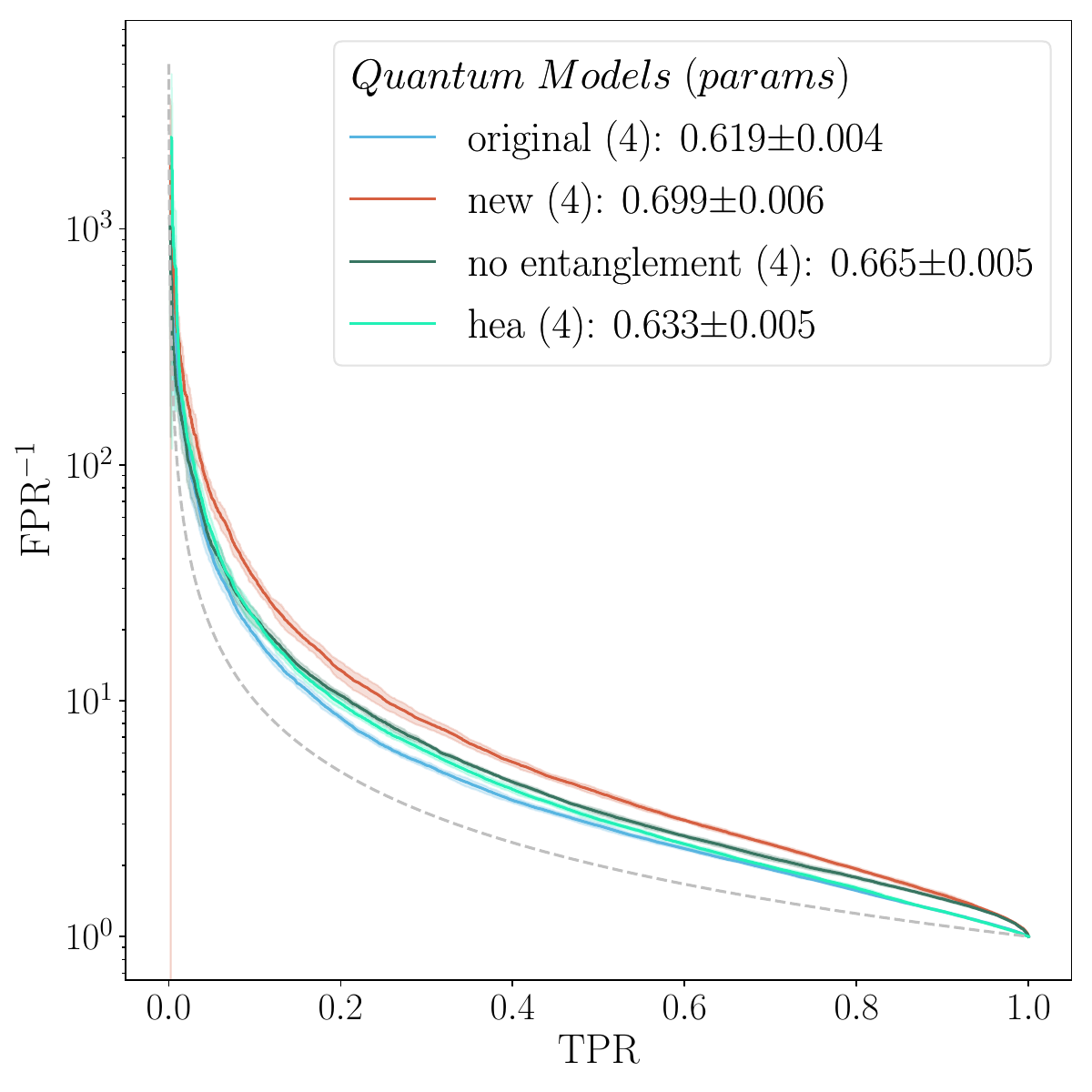}  \\
     (a) Narrow $G \rightarrow WW$, $8$ features &  $A \rightarrow HZ \rightarrow ZZZ$, $8$ features &  Wide $G \rightarrow WW$, $8$ features \\[6pt]
     \includegraphics[width=0.32\textwidth]{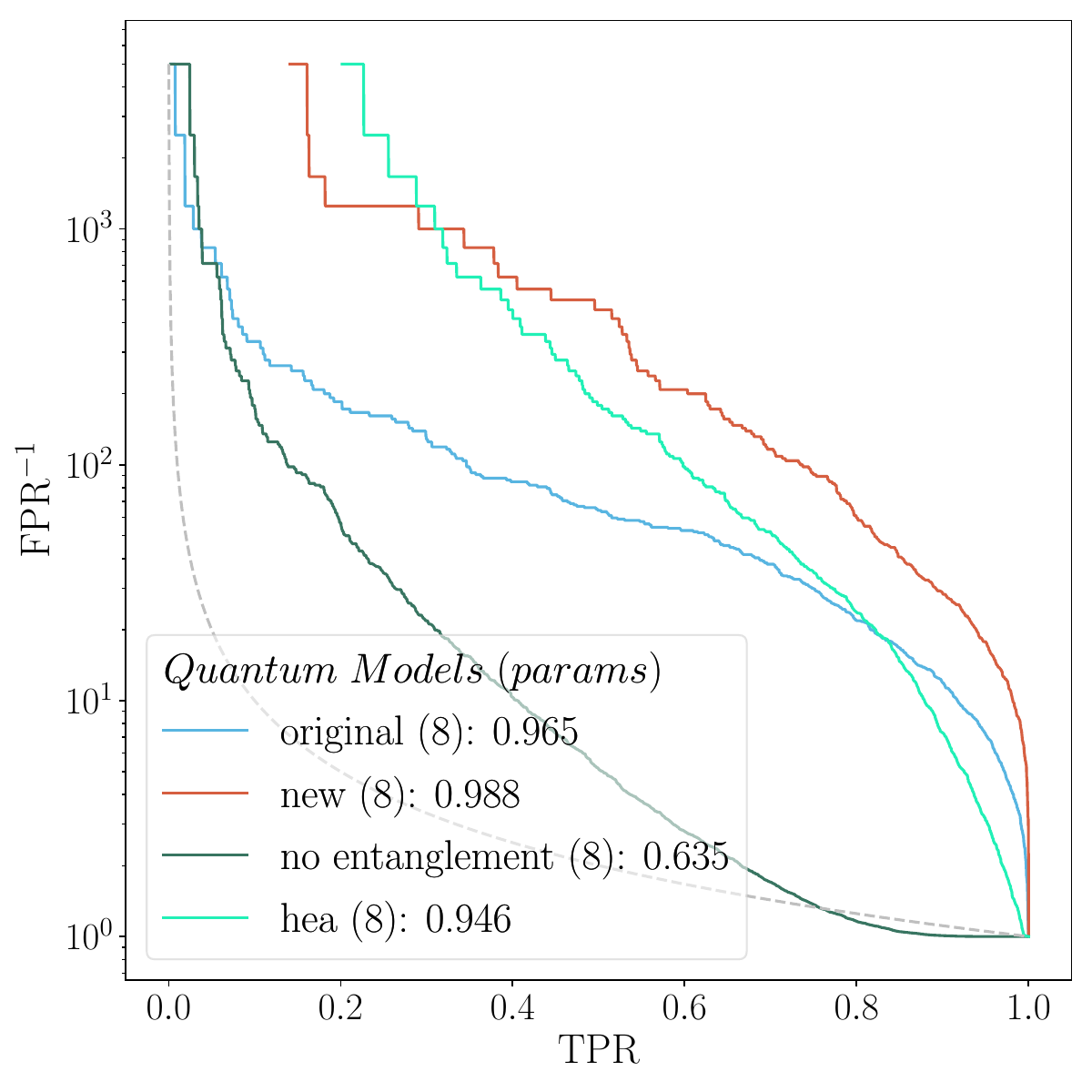} & \includegraphics[width=0.32\textwidth]{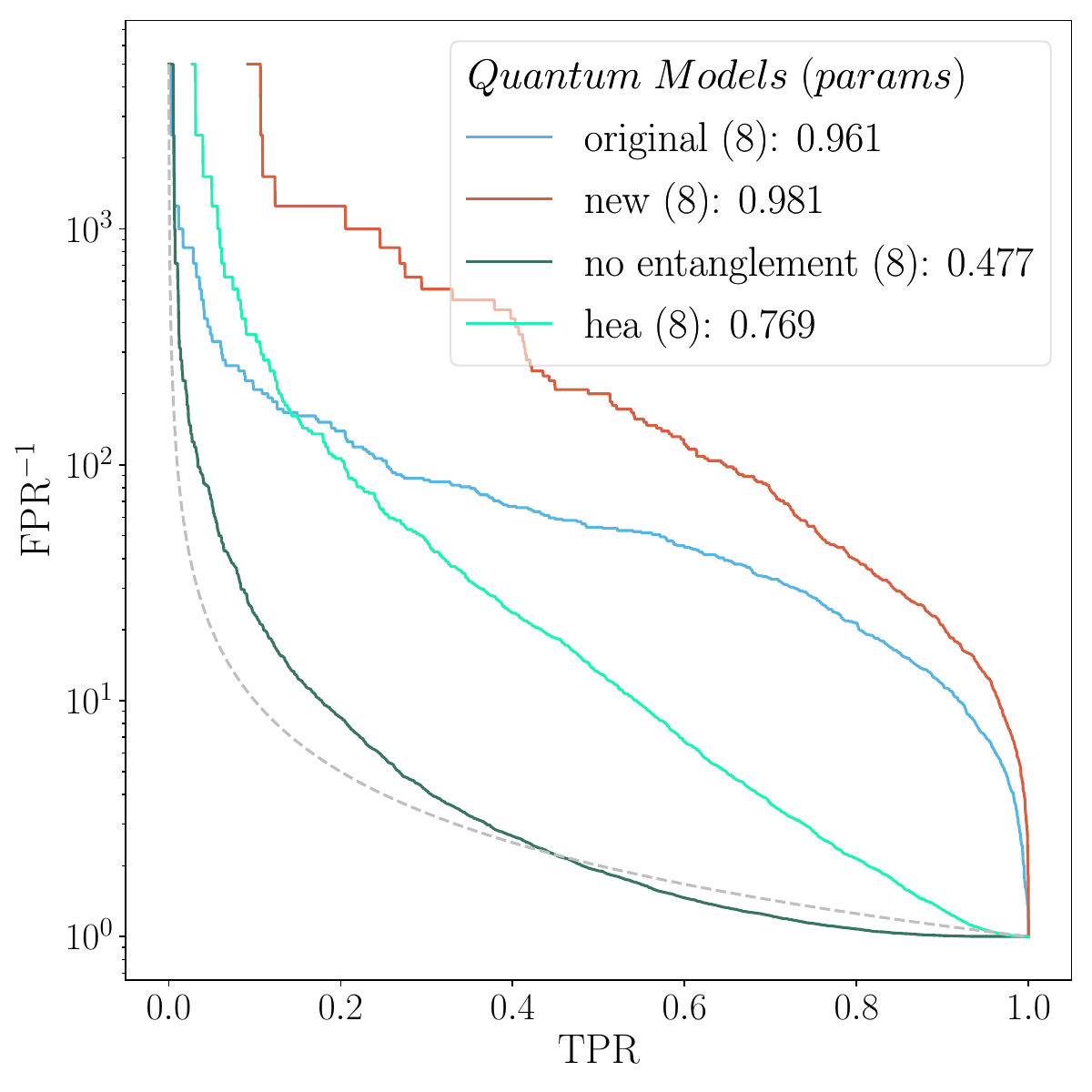} & \includegraphics[width=0.32\textwidth]{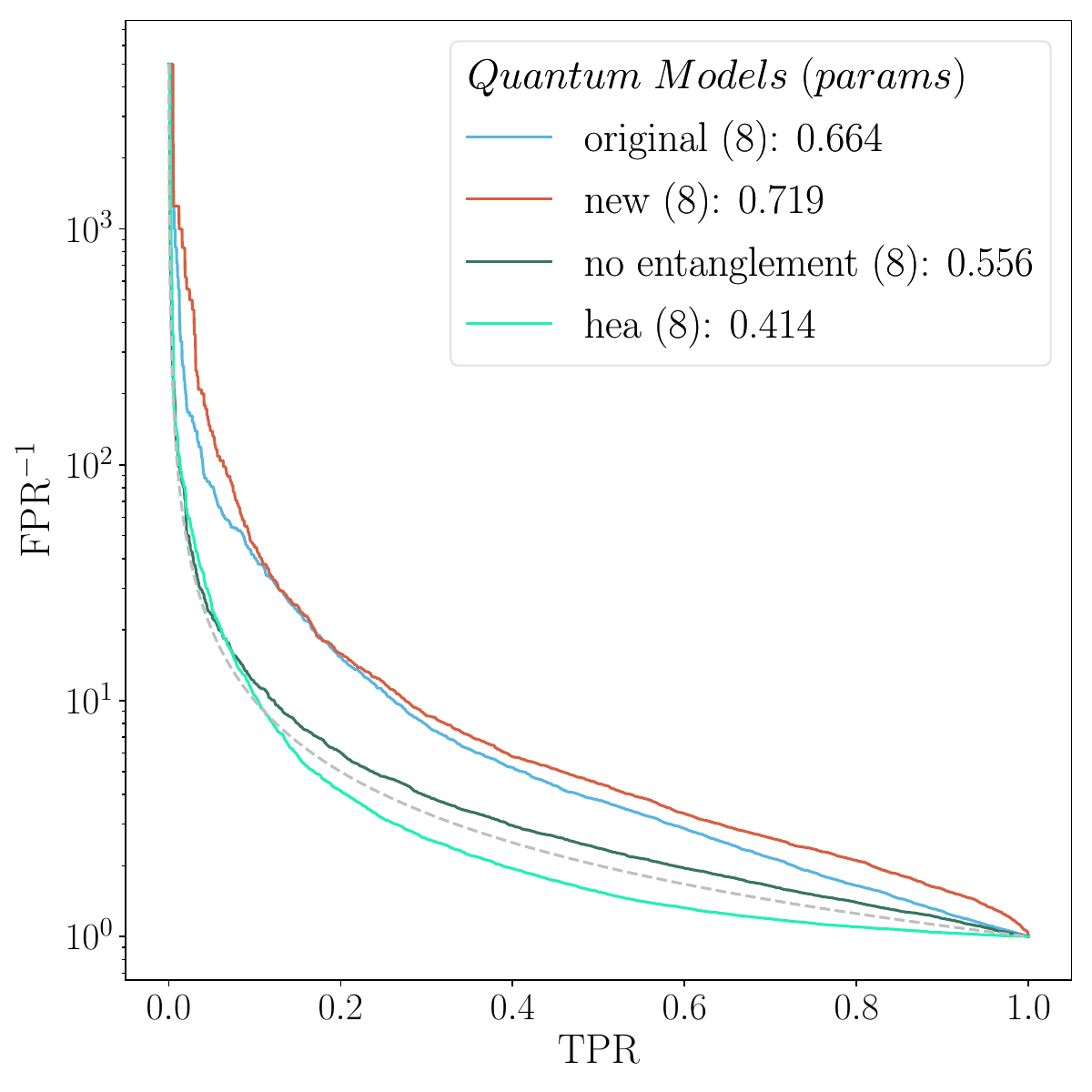} \\
     (b) Narrow $G \rightarrow WW$, $16$ features &  $A \rightarrow HZ \rightarrow ZZZ$, $16$ features &  Wide $G \rightarrow WW$, $16$ features \\[6pt]
\end{tabular}
\caption{Each subplot depicts the ROC curve for a collection of QAEs for various numbers of input features and BSM signals. Each column corresponds to identifying a different BSM signal, going from left to right we have the narrow graviton, scalar $ZZZ$ boson and wide graviton respectively. The QAEs depicted here each contain the same feature map seen in Figure \ref{circ:graviton_feat_map} but differing ansatz design. Row (a) eight input features with five folds of test data. Row (b) 16 input features with one fold of test data.}
\label{fig:qcd_roc_q}
\end{figure*}

\subsubsection{\label{sec:level5aa} Heavy Higgs}

We shall first consider the performance of both CAEs and QAEs to identify the heavy Higgs signal from the QCD background. The number of input features considered were four, six and eight, with the resulting ROC curves in Figure \ref{fig:higgs_roc}.\newline

All three quantum models perform similarly with four input features, achieving an average AUC score of $0.976 \pm 0.001$. The newly proposed ansatz, however, demonstrates its advantage as we increase the number of features considered. Its AUC score saturates at $0.983 \pm 0.001$ for four and eight features, while the other ansatz exhibits degradation.\newline 

For the original ansatz and HEA, the AUC scores drop significantly to $0.858 \pm 0.002$ and $0.753 \pm 0.005$ at six features and $0.212 \pm 0.002$ and $0.313 \pm 0.004$ at eight features, respectively. Scores below $0.5$ indicate performance worse than random guessing, highlighting the models' inability to learn the underlying data. This can be attributed to the higher loss reconstruction for background data than signal data, which is atypical. This inability of the original ansatz and HEA to learn the background data during training could be attributed to the solution space these ansatz possess not containing good approximations of the problem solution. Here, we see that despite each of the quantum models containing the same trainable rotation gates and thus equal expressibility, the differing entangling structures significantly impact final performance.\newline 

As illustrated in Figure \ref{fig:higgs_roc}, the newly introduced QAE ansatz outperforms all classical models at each input size. At four input features, this distinction is particularly evident when considering the shallow network but less so for the deep model. For models with six and eight features, the performance gap narrows between the CAEs and the new QAE but is still evident.\newline  

Finally, we can see that the number of parameters contained in the QAEs is far fewer than in the CAEs, even as problem size increases. This interesting point will be discussed in more detail in the context of the second data set.\newline   

\begin{figure*}[t!]
\begin{tabular}{ccc}
     \includegraphics[width=0.32\textwidth]{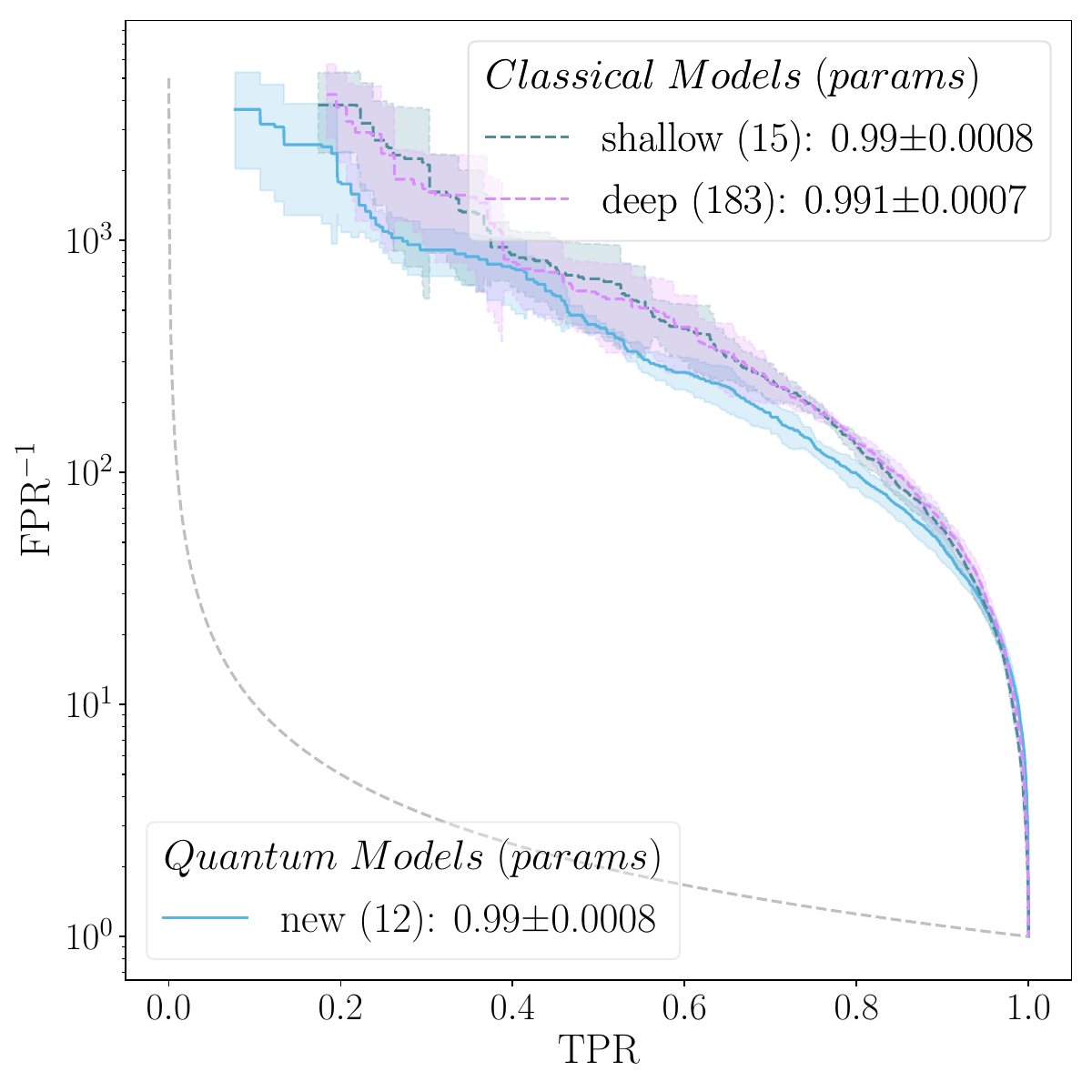}& \includegraphics[width=0.32\textwidth]{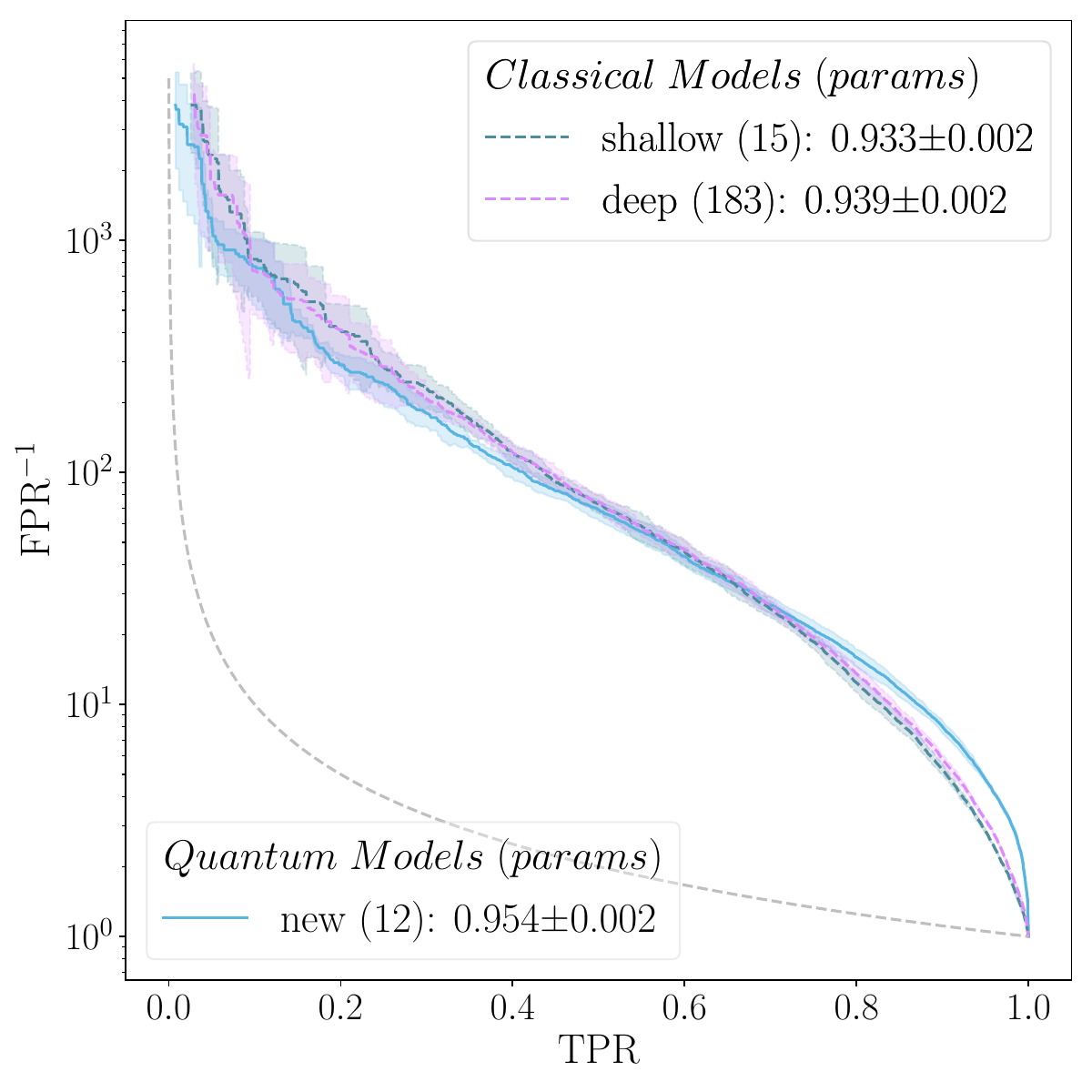} & \includegraphics[width=0.32\textwidth]{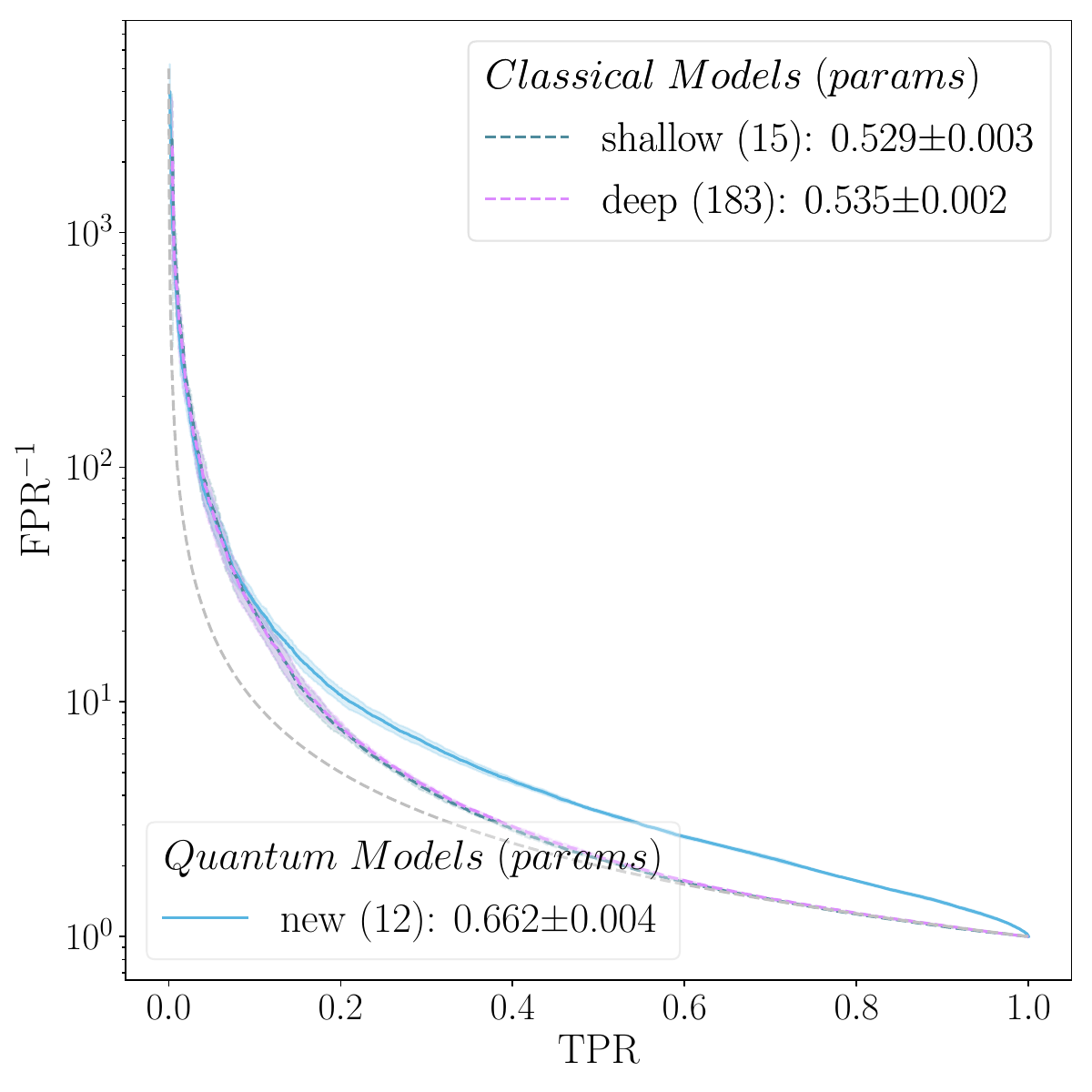}  \\
     (a) Narrow $G \rightarrow WW$, $8$ features &  $A \rightarrow HZ \rightarrow ZZZ$, $8$ features &  Wide $G \rightarrow WW$, $8$ features \\[6pt]
     \includegraphics[width=0.32\textwidth]{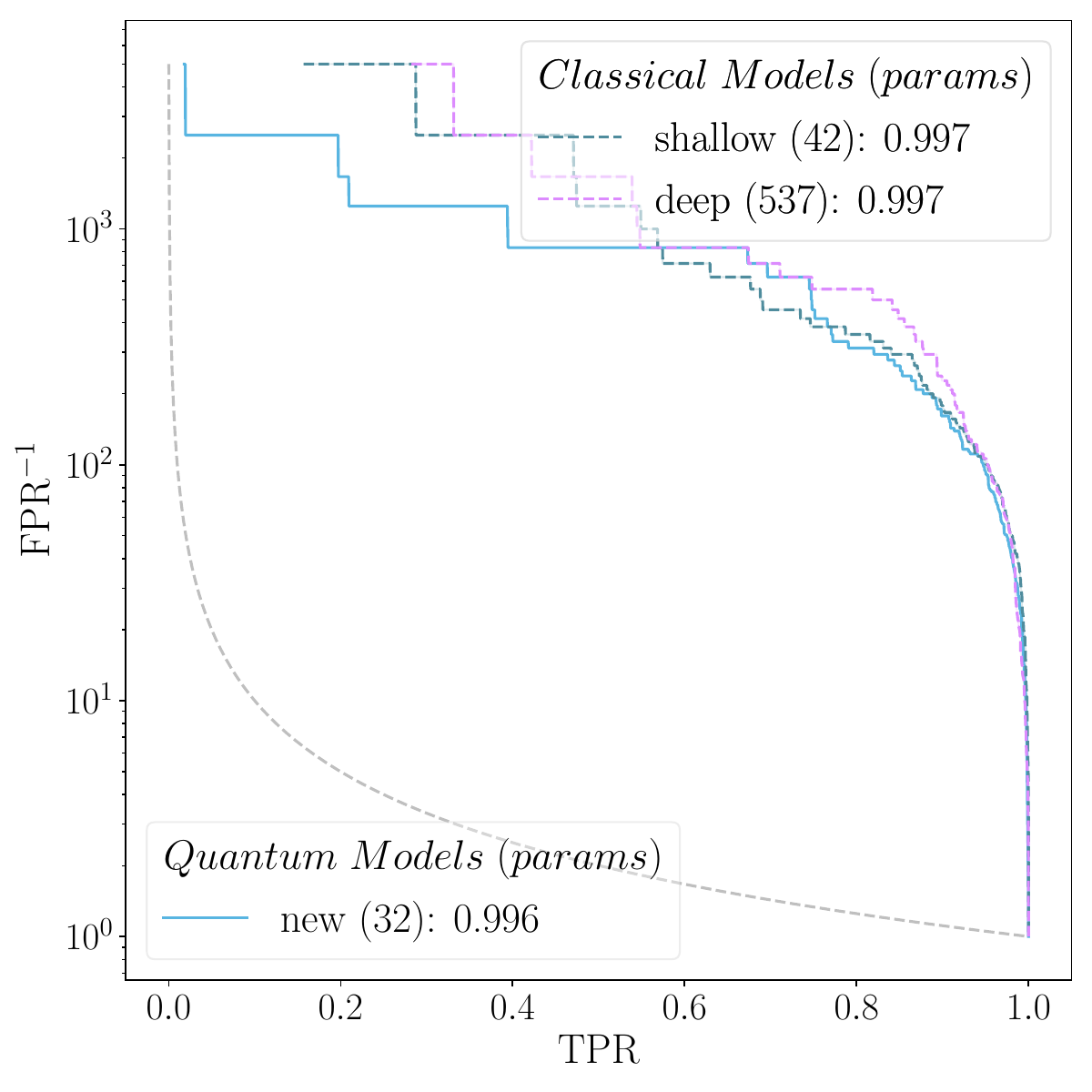} & \includegraphics[width=0.32\textwidth]{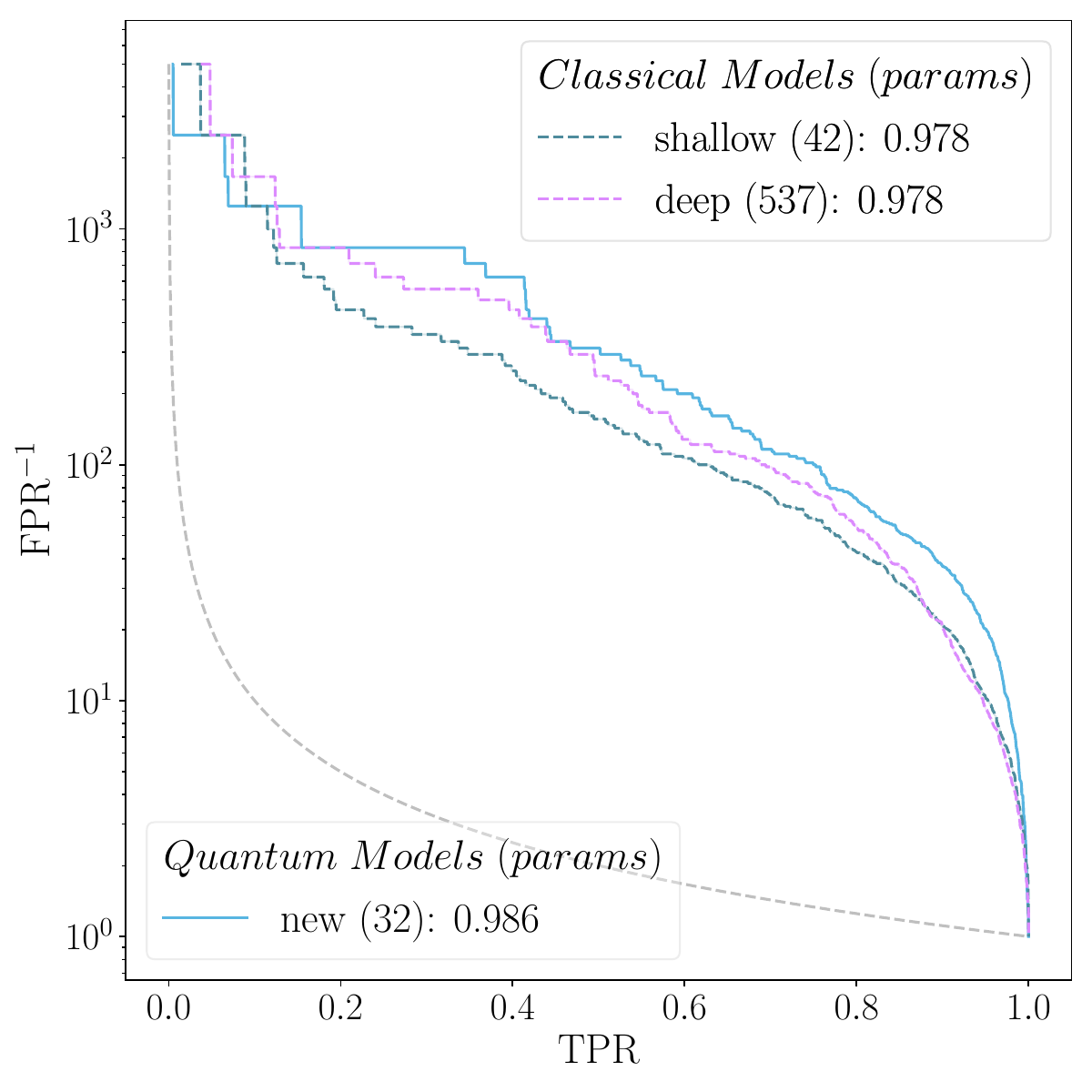} & \includegraphics[width=0.32\textwidth]{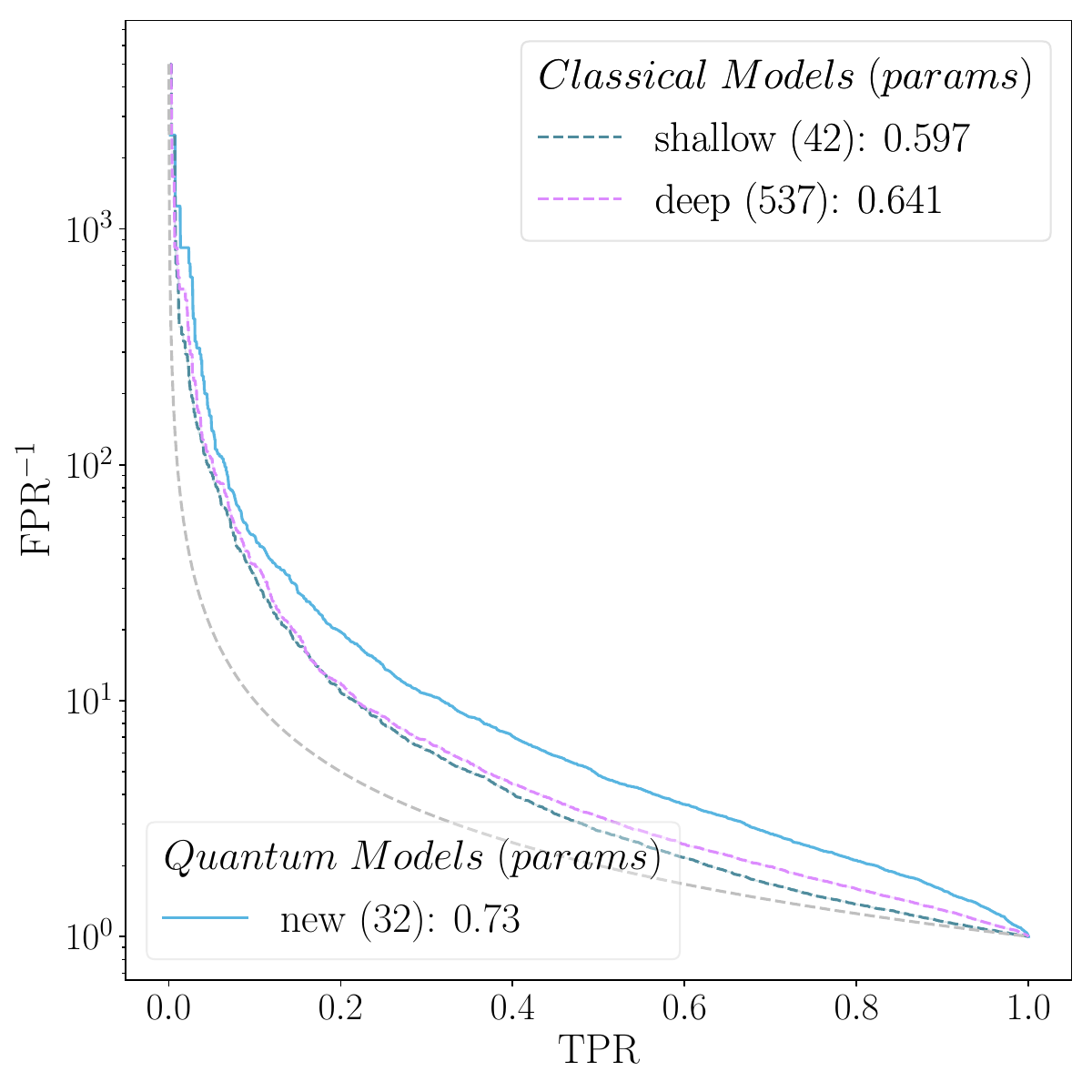} \\
     (b) Narrow $G \rightarrow WW$, $16$ features &  $A \rightarrow HZ \rightarrow ZZZ$, $16$ features &  Wide $G \rightarrow WW$, $16$ features \\[6pt]
\end{tabular}
\caption{Each subplot depicts the ROC curve for a collection of QAEs and CAEs for various numbers of input features and BSM signals. Each column corresponds to identifying a different BSM signal, going from left to right we have the narrow graviton, scalar $ZZZ$ boson and wide graviton respectively. The QAEs depicted here each contain the same feature map seen in Figure \ref{circ:graviton_feat_map} and ansatz design from Figure \ref{circ:new}. The CAEs come in two flavours - shallow and deep - some of which have been sparsified to reduce the number of parameters. Row (a) eight input features with five folds of test data. Row (b) 16 input features with one fold of test data.}
\label{fig:qcd_roc}
\end{figure*}

\subsubsection{\label{sec:level5ab} Gravitons \& scalar Bosons }

Following the heavy Higgs signal analysis, we compare the performance of the introduced ansatz for identifying graviton and scalar boson signals. Figure \ref{fig:qcd_roc_q} shows the performance of the quantum models. Once again, a separation between the new and old ansatz can be observed.\newline  

While Figure \ref{fig:qcd_roc_q} depicts single-layer ansatz, multi-layer architectures were used for comparison with the classical models, shown in Figure \ref{fig:qcd_roc}. In the top row (eight features), the new three-layer ansatz contains $12$ parameters similar to the $15$ in the shallow CAE, whereas for the deep CAE, $183$ parameters are present. Again, a performance separation is observed in favour of the new ansatz, except for identifying the narrow graviton, where all models perform equally.\newline

Similarly, the bottom row of Figure \ref{fig:qcd_roc} ($16$ features) shows the QAE with $48$ parameters just exceeding that of the $42$-parameter shallow CAE but remaining significantly less complex than the $810$-parameter deep network. While CAEs narrowly outperform the QAE in identifying the narrow graviton (AUC: $0.997$ vs $0.996$), the QAE surpasses the best-performing CAE for both the scalar boson (AUC: $0.986$ vs $0.978$) and broad graviton (AUC: $0.73$ vs $0.641$) cases.\newline   

These results suggest that QAEs can achieve superior discriminating power compared to classical models using fewer or comparable numbers of parameters. This empirically demonstrates the potential for QAEs to achieve higher performance with lower computational complexity.\newline

\subsection{\label{sec:level5b} Entanglement}

\begin{figure}
\includegraphics[width=0.5\textwidth]{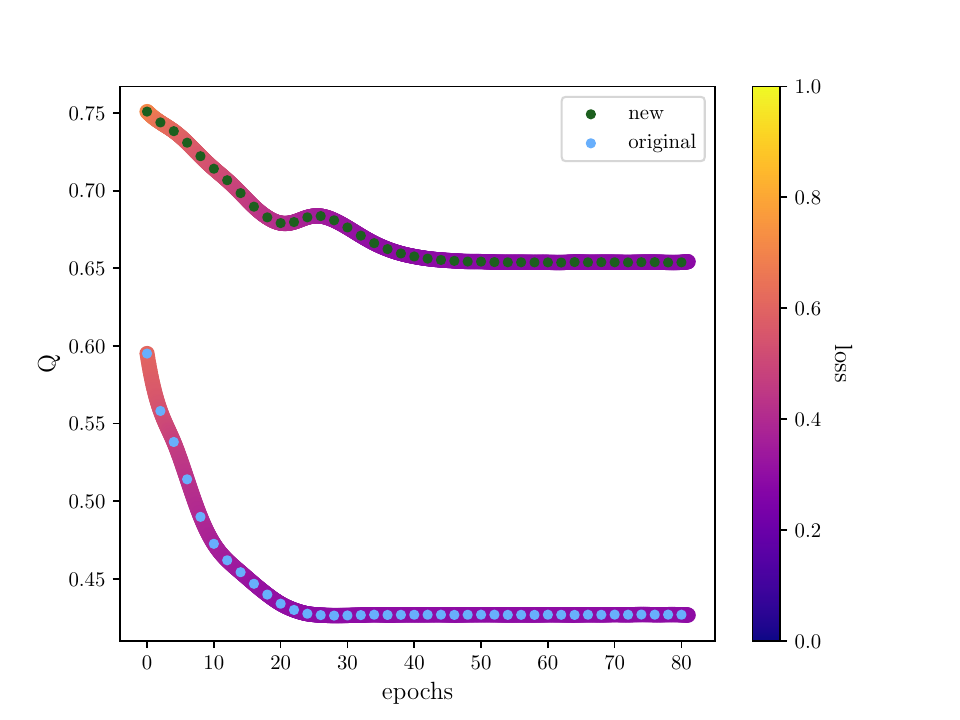}
\caption{An illustration of how global entanglement varies during training for two types of ansatz sampled every two epochs averaged over five folds of training. The corresponding loss during training was recorded in the same manner as entanglement and has been included via a color gradient, values for loss have been interpolated with a cubic spline. The validation set used for computing the above metrics was $5000$ background samples. Data describing QCD background processes from the `Graviton and Scalar Boson' dataset were used to construct this figure.}
\label{fig:dynamic_e}
\end{figure}

\begin{figure*}
\begin{tabular}{ccc}
     \includegraphics[width=0.3\textwidth]{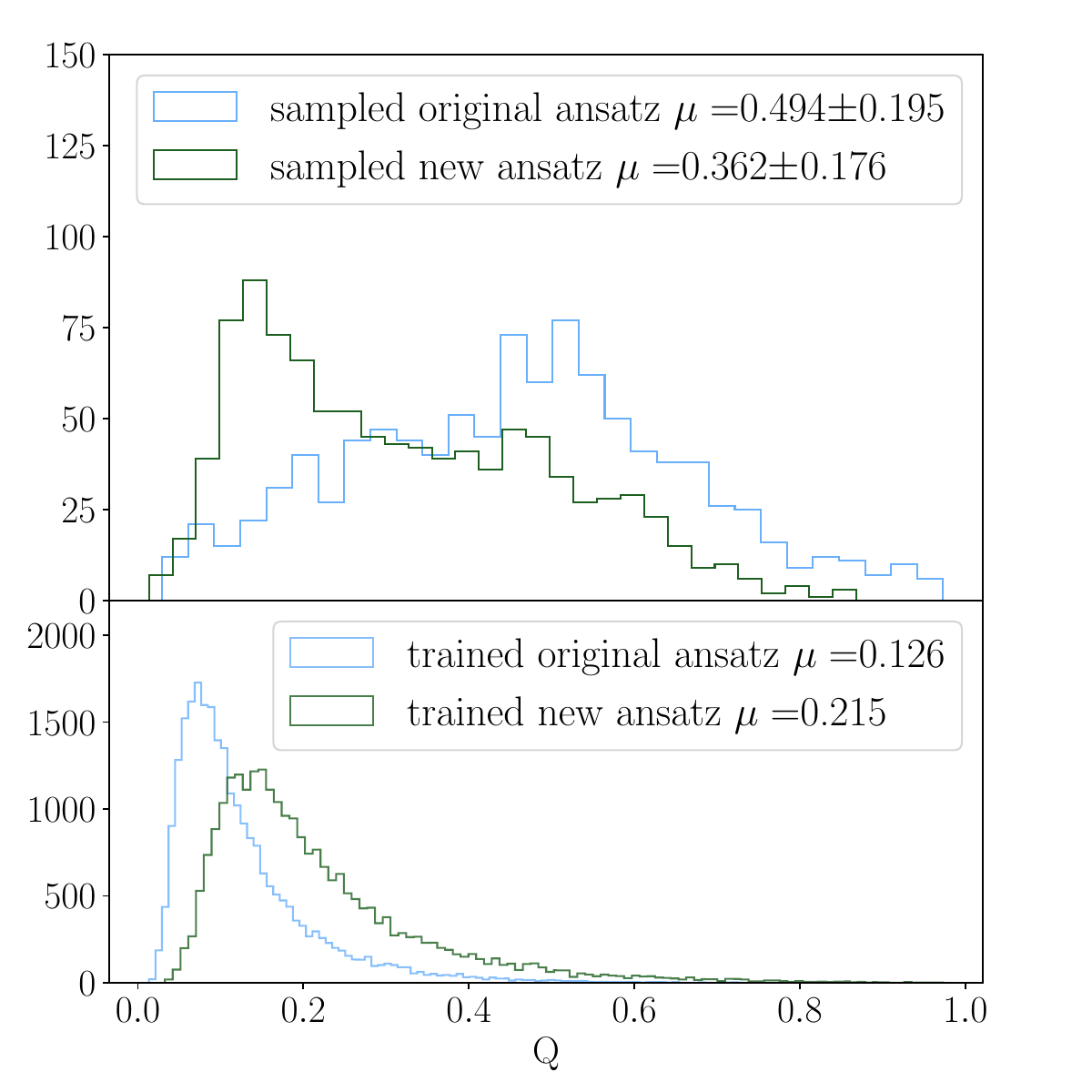}& \includegraphics[width=0.3\textwidth]{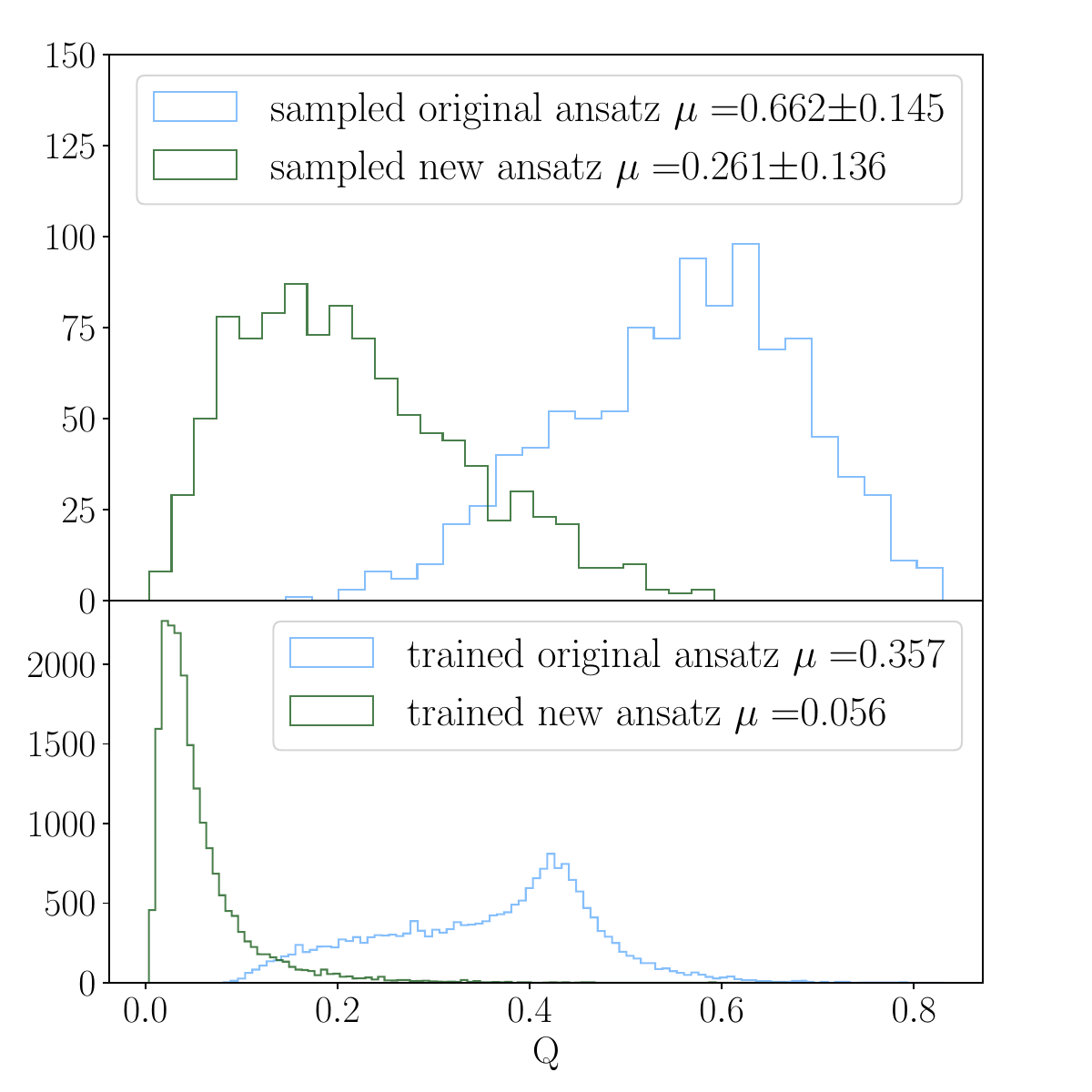} & \includegraphics[width=0.3\textwidth]{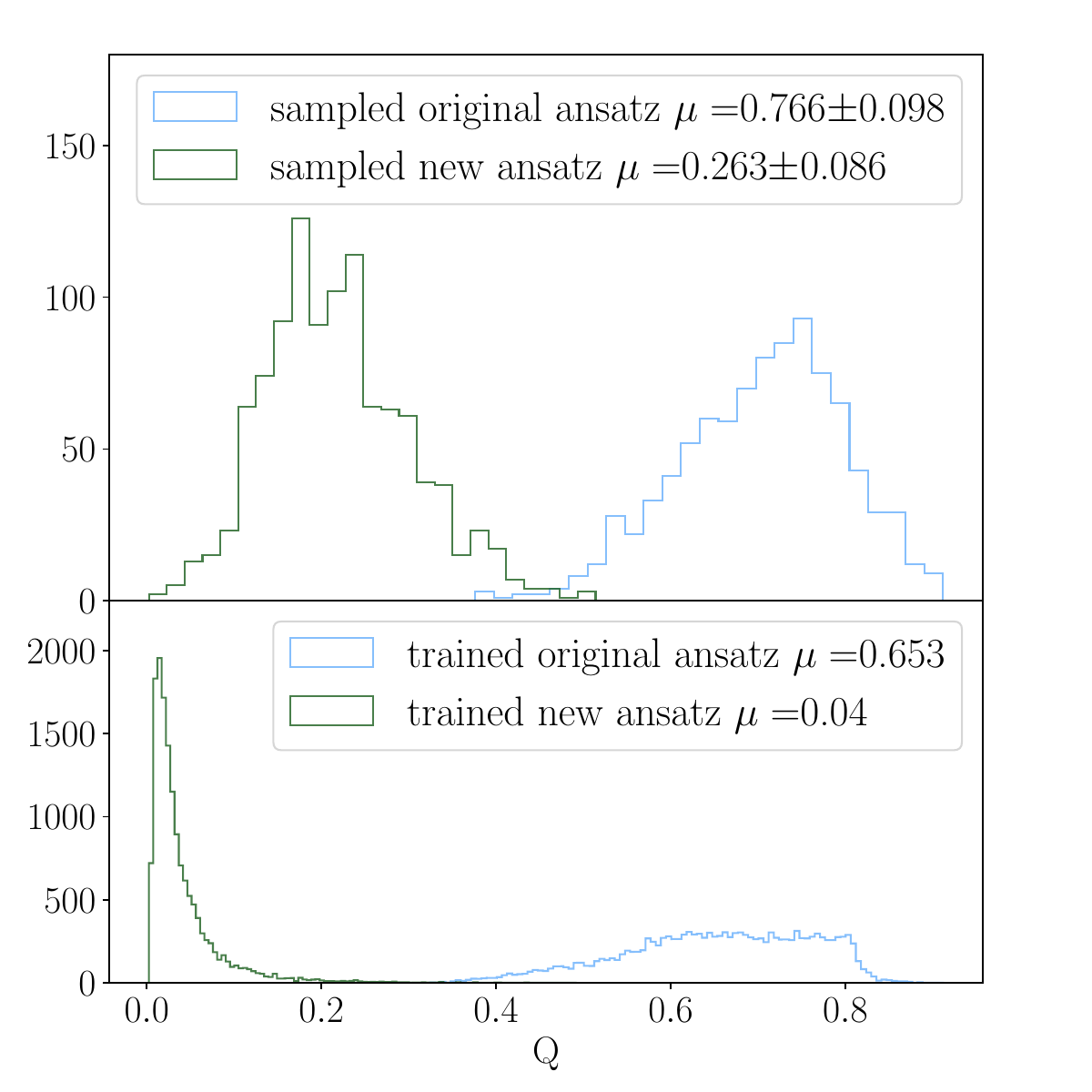}  \\
     (a) $4$ features &  $6$ features &  $8$ features \\[6pt]
     \includegraphics[width=0.3\textwidth]{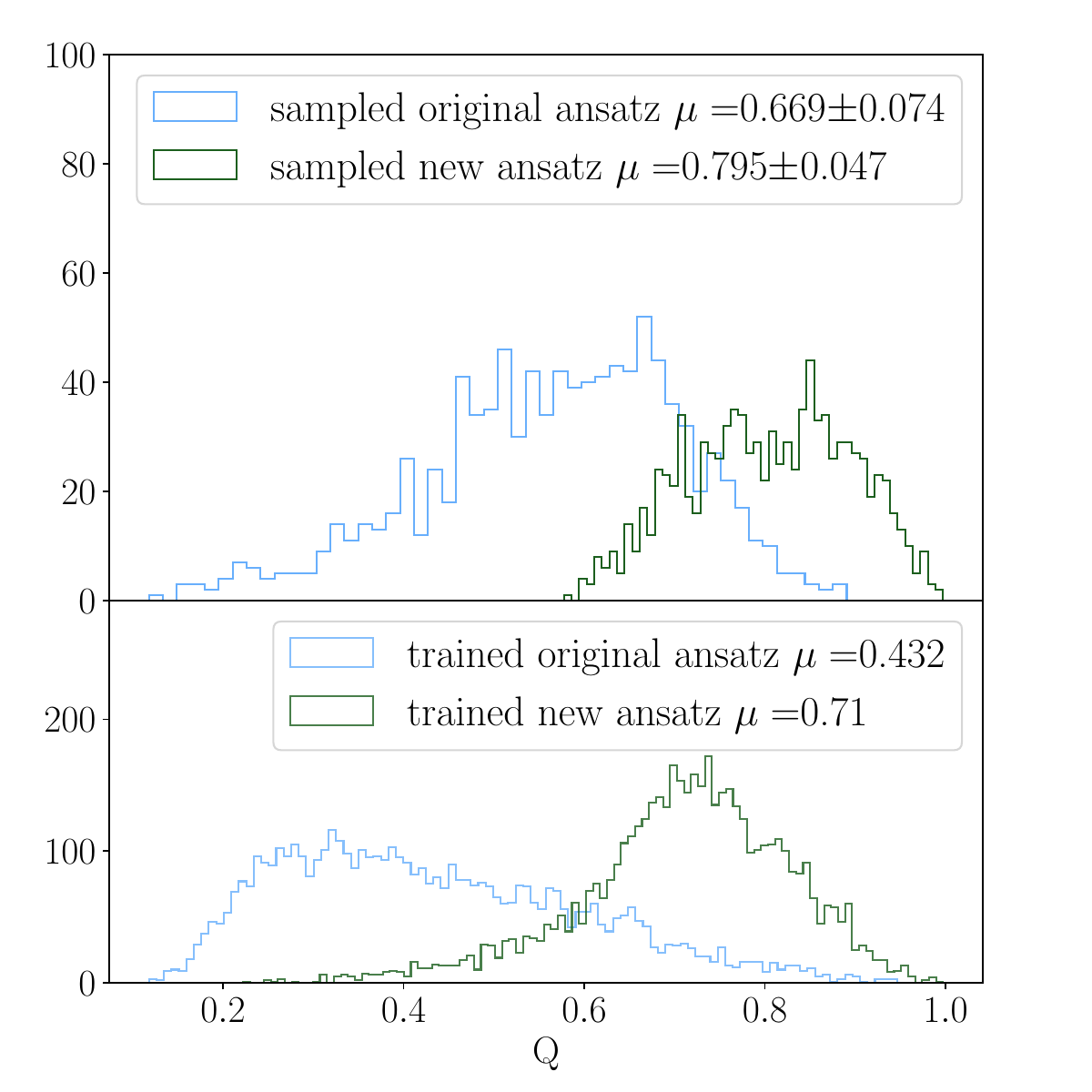} & \includegraphics[width=0.3\textwidth]{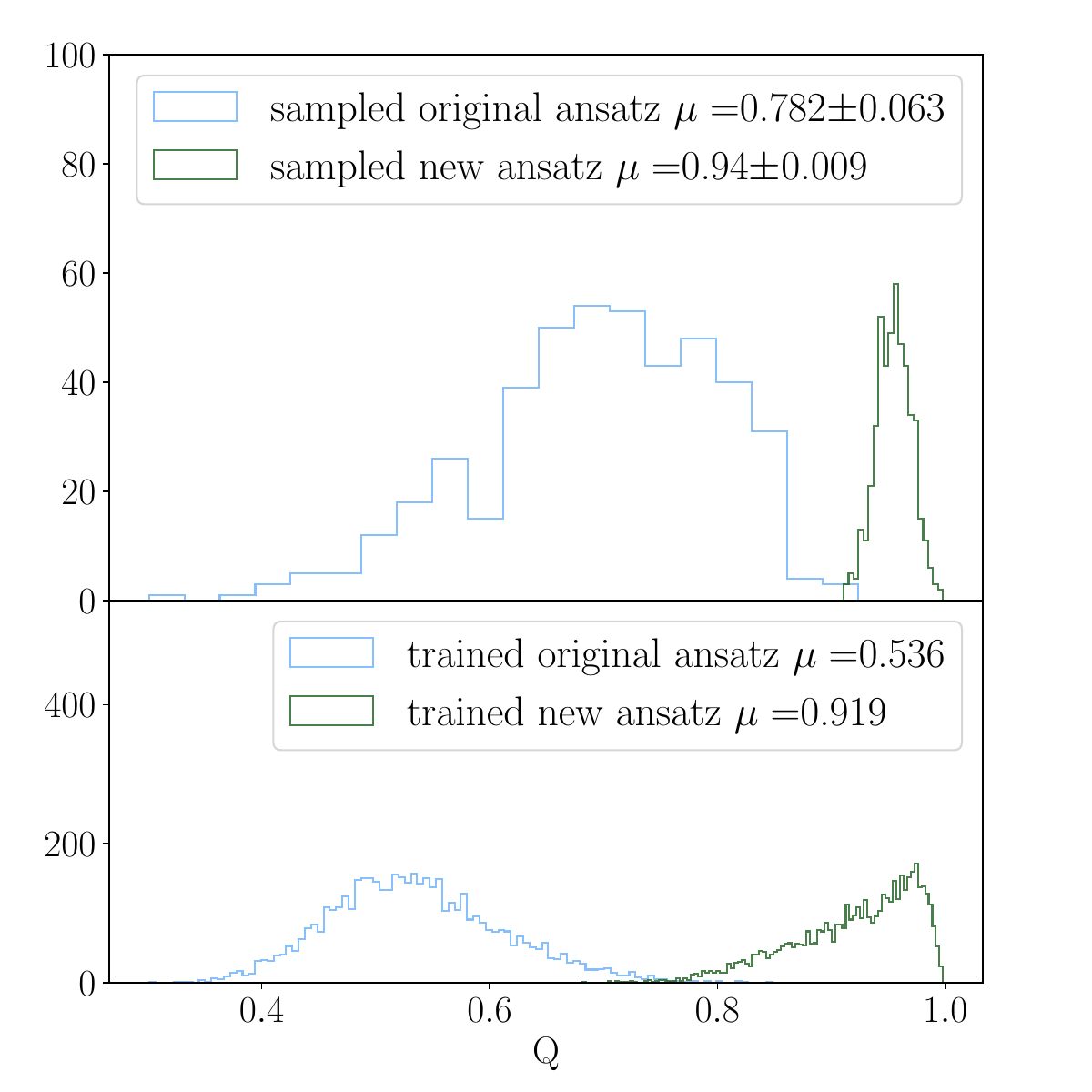} & 
     \\
     (b) $8$ features &  $16$ features \\[6pt]
\end{tabular}
\caption{Histograms showing the amount of global entanglement generated by circuits using the Meyer-Wallach entanglement measure $Q$. Each subplot consists of two histograms, the lower histogram is the distribution of entanglement for a circuit over a set of background samples $x$, with a set of trained parameters $\theta_{trained}$, we can define the distribution via $Q(|\psi(x)|_{\theta_{trained}}\rangle)$. The upper histogram depicts a distribution given by  $\mathbb{E}[Q(|\psi(\theta, x)\rangle$], where $\theta$ is sampled from the uniform distribution. This provides us a distribution on the expectation values of $Q$ the circuit of interest could take. Row (a) Background samples are taken from the `heavy Higgs' dataset, the circuit sizes considered are four, six and eight qubits going from left to right respectively. Row (b) Background samples are taken from the `scalar boson and graviton' dataset, the circuit sizes considered are four and eight qubits going from left to right respectively.} 
\label{fig:entanglement}
\end{figure*}

We investigate the behaviour of global entanglement for a QAE during training and analyse how this entanglement evolves throughout the process, as illustrated in Figure \ref{fig:dynamic_e}. For every two epochs, the global entanglement of the original and the new ansatz was calculated over three training folds using a validation set of $5000$ samples. The observed trend is a decrease in global entanglement alongside decreasing loss function values, suggesting convergence to a state with less entanglement than is, on average, achieved by the ansatz.\newline 

To gain further insights, we present histograms of the global entanglement for circuits with randomly sampled parameters in Figure \ref{fig:entanglement}. The top row shows circuits trained on background samples from the `heavy Higgs' dataset, while the bottom row shows those trained on the background samples from the `scalar boson and graviton' dataset. Each subplot consists of two histograms:

\begin{itemize}
    \item \textbf{Lower Histograms}: Represents the distribution of global entanglement over the dataset for a set of parameters found after training, denoted as $Q(|\psi(x)|_{\theta_{trained}}\rangle)$. The mean of this distribution is equivalent to equation \ref{eq:entanglement_data_mean}.
    \item \textbf{Upper Histograms}: Shows the distribution of global entanglement for the dataset across multiple randomly sampled parameter sets, represented by $\dfrac{1}{N}\sum_i^NQ(|\psi(x_i,\theta)\rangle)$. The mean of this distribution is equivalent to equation \ref{eq:entanglement_data_param_mean}.
\end{itemize}

Across all plots in Figure \ref{fig:entanglement}, a consistent trend emerges. The final, optimized parameters consistently generate a global entanglement with a lower expected value compared to the bulk of the distribution arising from uniformly sampled parameters. In the cases shown for the new ansatz, all but the first plot, which considers four input features from the `heavy Higgs' dataset, show $\mathbb{E}[Q(|\psi(x)|_{\theta_{trained}}\rangle)]$ to be over $1\sigma$ from $\mathbb{E}[\dfrac{1}{N}\sum_i^NQ(|\psi(x_i,\theta)\rangle)]$. This suggests that the QAE exhibits a preference for lower entanglement states.\newline 

One potential explanation for this preference is that high entanglement levels might lead to information scrambling, making lower entanglement states more favourable. Previous work explored this idea, showing that too much entanglement can render quantum states useless for quantum computation \cite{PhysRevLett.102.190501}.\newline 

Furthermore, focusing on the top row of Figure \ref{fig:entanglement}, we observe that the entanglement generated by the new ansatz remains localized with increasing problem size. In contrast, the original ansatz exhibits a flattening in its distribution with increasing problem size, coinciding with its observed decrease in performance.\newline 

A final observation is that the entanglement decreases for the new ansatz as the problem size increases when considering the `heavy Higgs' dataset in the top row of Figure \ref{fig:entanglement}. Meanwhile, for the new ansatz, entanglement increases when considering the `scalar boson and graviton' dataset in the bottom row of Figure \ref{fig:entanglement}. The exact reasons for these trends may depend on factors such as entanglement in the feature map or the data itself. No entanglement was present in the feature map for the `heavy Higgs' dataset, but for the `scalar boson and graviton' dataset, there was. This could be a topic for further exploration since entanglement growth is necessary in order to avoid dequantization \cite{eisert2013entanglement}. 

\subsection{\label{sec:level5c} Magic}
\begin{figure}
\includegraphics[width=0.5\textwidth]{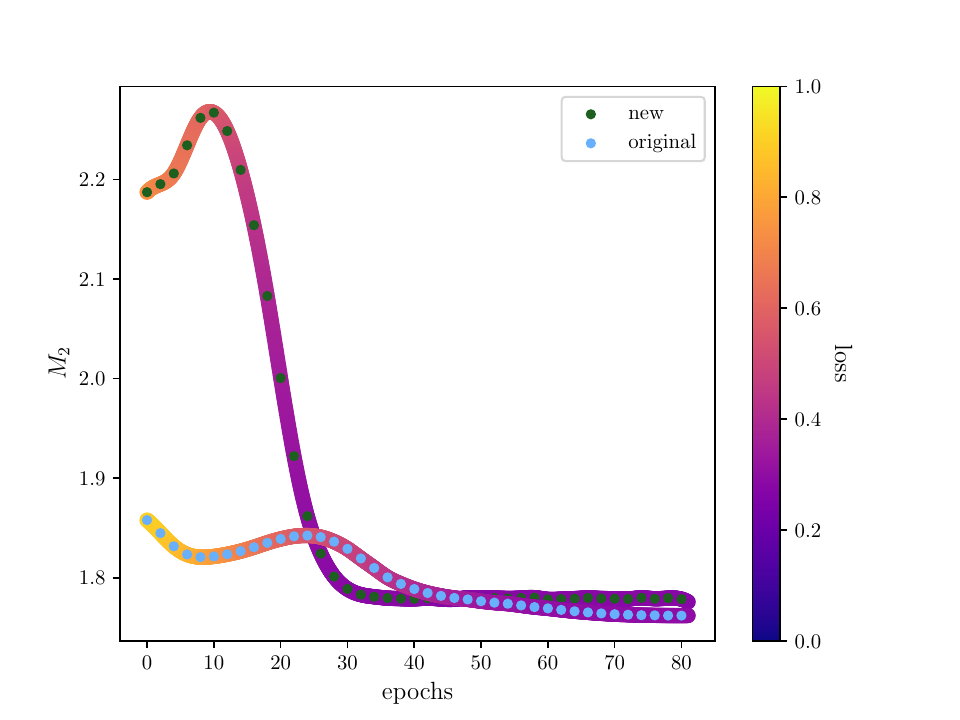}
\caption{An illustration of how magic varies during training for two types of ansatz sampled every two epochs averaged over five folds of training. The corresponding loss during training was recorded in the same manner as magic and has been included via a color gradient, values for loss have been interpolated with a cubic spline. The validation set used for computing the above metrics was $1000$ background samples. Data describing QCD background processes from the `Graviton and Scalar Boson' dataset were used to construct this figure.}
\label{fig:dynamic_m}
\end{figure}

\begin{figure}
\begin{tabular}{c}
     \includegraphics[width=0.4\textwidth]{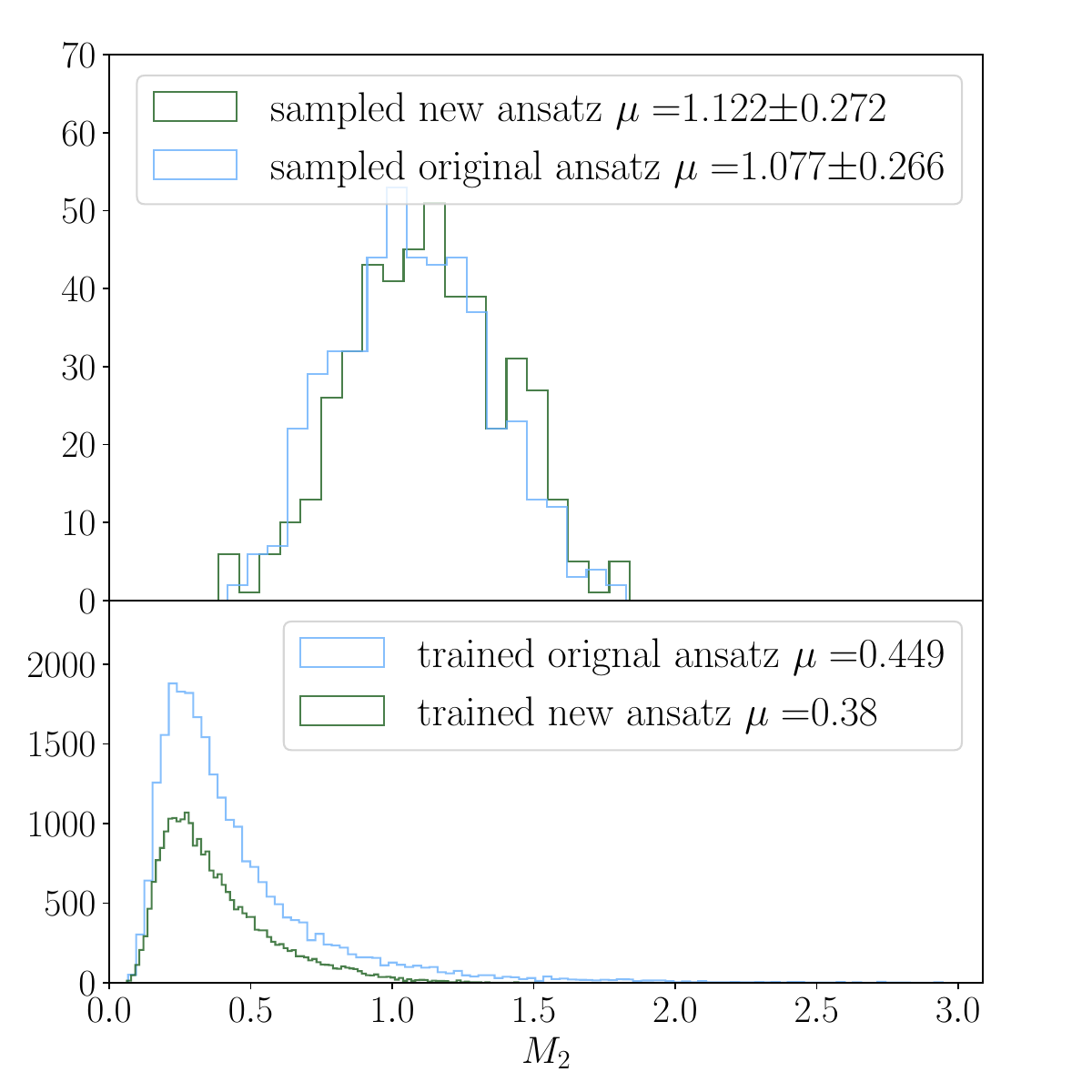} \\
     (a) \\[6pt]
     \includegraphics[width=0.4\textwidth]{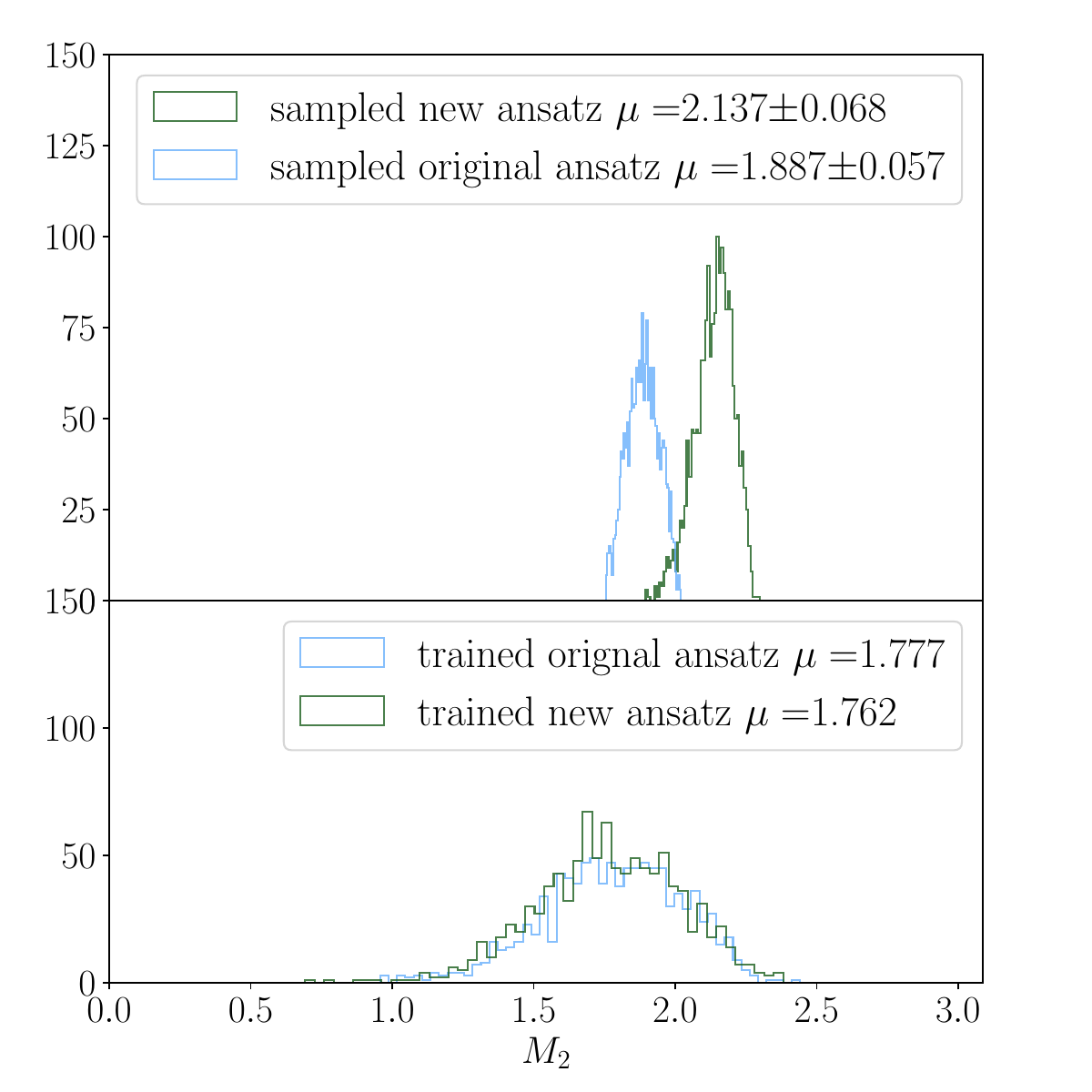} \\
     (b) \\[6pt]
\end{tabular}
\caption{A collection of histograms depicting the amount of magic generated by circuits using the stabilizer R\'enyi entropy $M_2$. Each subplot consists of two histograms, the lower histogram is the distribution of magic for a circuit over a set of background samples $x$, with a set of trained parameters $\theta_{trained}$, we can define the distribution via $M_2(|\psi(x)|_{\theta_{trained}}\rangle)$. The upper histogram depicts a distribution given by  $\mathbb{E}[M_2(|\psi(\theta, x)\rangle$)], where $\theta$ is sampled from the uniform distribution. This provides us a distribution on the expectation values of $M_2$ the circuit of interest could take. Row (a) Background samples are taken from the `heavy Higgs' dataset, for a circuit size of four qubits. Row (b) Background samples are taken from the `scalar boson and graviton' dataset for a circuit size of four qubits.}
\label{fig:magic_dist}
\end{figure}

Following our analysis of entanglement, we adopt a similar approach to investigate the behaviour of magic during training. Figure \ref{fig:dynamic_m} shows the magic generated by the original and new ansatz, recorded every two epochs over three training folds with a validation set of $1000$. Similar to the observed trend for global entanglement, we see a decrease in magic alongside decreasing loss function values, suggesting convergence towards lower magic states.\newline  

Analogous to the entanglement analysis, Figure \ref{fig:magic_dist} presents histograms of magic with uniformly randomly sampled parameters. The x-axis limits on the histograms are purposely set to be the minimum and maximum attainable magic for the system being considered. The subplots depict:

\begin{itemize}
    \item \textbf{Lower Histograms}: Represents the distribution of magic over the dataset for a set of parameters found after training, denoted as $M_2(|\psi(x)|_{\theta_{trained}}\rangle)$. The mean of this distribution is equivalent to \ref{eq:magic_data_mean}.
    \item \textbf{Upper Histograms}: Depicts the distribution of magic for the dataset across multiple randomly sampled parameter sets, represented by $\dfrac{1}{N}\sum_i^NM_2(|\psi(x_i,\theta)\rangle)$. The mean of this distribution is equivalent to equation \ref{eq:magic_data_param_mean}.
\end{itemize}

As evident in Figure \ref{fig:magic_dist}, the magic generated by circuits with optimized parameters consistently falls to the left of the mean compared to randomly sampled circuits. This effect is more pronounced for the `heavy Higgs' dataset than the `scalar boson and gravitons' dataset. Previous work has identified excess amounts of magic as useless for obtaining quantum advantage in the context of quantum many-body systems \cite{PRXQuantum.3.020333}. For the circuits considered in this work, states with maximal magic are not produced. It is interesting that, in fact, the most valuable states are those with less magic. Since this is one of the first studies into entanglement and magic simultaneously for PQCs, further research is required to explore the interplay between these quantities and the usefulness of states with certain levels of each quantity.\newline 

\section{\label{sec:level6}Conclusion and Outlook}

In conclusion, we investigated and improved upon two existing studies in anomaly detection in high energy physics by proposing a novel QAE ansatz, considering up to $16$ features. The new QAE ansatz demonstrated a performance advantage in both datasets compared to previously considered ansatz. This may suggest that the ansatz is applicable across a variety of datasets. This study addresses larger, more relevant problem sizes in the context of particle physics. However, future studies using even larger problem sizes will be key to verifying that the trend for QML models matching or beating their classical counterparts using fewer parameters continues to hold.\newline

Our study highlights the importance of designing ansatz that respect the problem's inherent structure. The newly developed QAE, which incorporates this principle, outperformed previously considered ansatz despite all three sharing the same level of expressibility. This finding suggests that the specific entangling structure can significantly impact performance. In particular, it allows for scaling to more qubits, which is not observed for previously used models.\newline 

We observed a clear separation in performance between the QAE and CAEs, with the QAE often requiring significantly fewer parameters to achieve superior performance in both datasets. Additionally, we explored the behaviour of global entanglement and magic generated by quantum circuits during training, demonstrating a convergence towards values smaller than the expectation value for the circuit family in both metrics.\newline

Further studies could investigate the scaling of entanglement and magic with increasing problem size to determine if optimised circuits possess sufficient levels of the two to resist dequantisation methods such as tensor networks \cite{Shin:2023fzq, PhysRevLett.129.090502, cerezo2024does}. Moreover, we believe it is important to understand the origin behind this observed preference for lower entanglement and magic during optimisation. This could involve analysing the relationships that entanglement and magic may have with the feature map used.\newline 

\begin{acknowledgments}
We acknowledge funding from the STFC. CD acknowledges CDT-DIS grants for their support of this work. SM is funded by grants from the Royal Society. 
\end{acknowledgments}

\FloatBarrier
\bibliography{apssamp}

\FloatBarrier
\newpage
\appendix

\section{\label{appx:pseudocode}New ansatz pseudocode}

The accompanying pseudocode for the newly proposed ansatz for the QAE. 

\begin{algorithm}
    \caption{New ansatz pseudo-code}\label{alg:new}
    \begin{algorithmic}
        \State \textbf{Input:} N (number of qubits), L (latent space), $\ell$ (layers)
        \State T = N - L
        \For{$i = 0 \to \ell - 1$}
            \State counter\_1 $\gets$ 0
            \For{$t = 0 \to T - 1$}
                \If{counter\_1 $\geq$ L}
                    \State counter\_1 $\gets$ 0
                \EndIf
                \State control $\gets N - 1 - t$
                \State target $\gets$ counter\_1
                \State CNOT(control, target)
                \State counter\_1 $\gets$ counter\_1 + 1
            \EndFor
        
            \For{$j = 0 \to N - 1$}
                \State $R_y(\theta_{j, i})$
            \EndFor
        
            \If{$i = \ell - 1$}
                \State counter\_2 $\gets$ 0 
                \For{$t = 0 \to T - 1$}
                    \State target $\gets N - 1 - t$
                    \If{target $<$ counter\_2}
                        \State counter\_2 $\gets$ 0
                    \EndIf
                    \State control $\gets$ counter\_2
                    \State CNOT(control, target)
                    \State counter\_2 $\gets$ counter\_2 + 1
                \EndFor
            \EndIf
        \EndFor
    \end{algorithmic}
\end{algorithm}

\section{\label{appx:model_details}Classical and quantum autoencoder details}

The tables presented here detail the hyperparameters of the QAEs and CAEs used in the above studies. Not listed in the tables for the CAEs are the activation functions used, these were leaky-relu functions for all hidden layers bar those connecting to the latent space which made use of the linear activation function. A learning rate of $0.001$ along with the Adam optimizer was used for the CAEs. For the QAEs the Adam optimizer was also used but with a learning rate of $0.005$.\newline  

\begin{table}[htbp!]
\vspace{0.3cm}
\begin{ruledtabular}
\begin{tabular}{ccccc}
Input size &Batch Size&Hidden Layers&Epochs&Sparsity\\
\hline
4 (shallow) & 500 & [4,3,1] & 500 & 0.400\\
4 (deep) & 500 & [4,4,3,2,1] & 500 & None\\
6 (shallow) & 500 & [3,3,2] & 500 & 0.305\\
6 (deep) & 500 & [8, 8, 8, 5, 4] & 500 & None\\
8 (shallow) & 500 & [5, 4] & 500 & 0.553\\
8 (deep) & 500 & [32, 32, 8, 4] & 500 & 0.750\\
\end{tabular}
\end{ruledtabular}
\caption{Details of the classical autoencoders used for identifying the heavy Higgs signal depicted in Figure \ref{fig:higgs_roc}.}
\label{table:Higgs_CAE}

\vspace{0.3cm}
\begin{ruledtabular}
\begin{tabular}{cccccc}
Input size &Batch Size&Layers&Epochs&Latent Space\\
\hline
4 & 50 & 1 & 100 & 1 \\
6 & 50 & 1 & 100 & 3 \\
8 & 50 & 1 & 100 & 4 \\
\end{tabular}
\end{ruledtabular}
\caption{Details of the quantum autoencoders used for identifying the heavy Higgs depicted in Figure \ref{fig:higgs_roc}.}
\label{table:Higgs_QAE}

\vspace{0.3cm}
\begin{ruledtabular}
\begin{tabular}{cccccc}
Input size &Batch Size&Layers&Epochs&Latent Space\\
\hline
8 & 50 & 1 & 100 & 1 \\
16 & 50 & 1 & 100 & 3 \\
\end{tabular}
\end{ruledtabular}
\caption{Details of the quantum autoencoders used for identifying a scalar boson and gravitons depicted in Figure \ref{fig:qcd_roc_q}.}
\label{table:Boson_QAE}

\vspace{0.3cm}
\begin{ruledtabular}
\begin{tabular}{ccccc}
Input size &Batch Size&Hidden Layers&Epochs&Sparsity\\
\hline
8 (shallow) & 500 & [1] & 500 & 0.731\\
8 (deep) & 500 & [5, 4, 3, 1] & 500 & None\\
16 (shallow) & 50 & [2] & 500 & 0.652\\
16 (deep) & 500 & [8, 6, 5, 5, 1] & 500 & None\\
\end{tabular}
\end{ruledtabular}
\caption{Details of the classical autoencoder used for identifying a scalar boson and gravitons depicted in Figure \ref{fig:qcd_roc}.}

\vspace{0.3cm}
\begin{ruledtabular}
\begin{tabular}{cccccc}
Input size &Batch Size&Layers&Epochs&Latent Space\\
\hline
8 & 50 & 3 & 40 & 2\\
16 & 50 & 6 & 60 & 5\\
\end{tabular}
\end{ruledtabular}
\caption{Details of the quantum autoencoders used for identifying a scalar boson and gravitons depicted in Figure \ref{fig:qcd_roc}.}
\end{table}

\section{\label{appx:grid_search}Grid search}
\begin{table}[H]
\vspace{0.3cm}
\begin{ruledtabular}
\begin{tabular}{cc}
Hyperparameter & Values\\
\hline
Batch Sizes & [50,500,1000] \\
Hidden Layers & range[1,5] \\
Neurons & range[1,32] \\
Latent Space & range[1,4] \\
Prune & range[0,1] \\
\end{tabular}
\end{ruledtabular}
\caption{Details of the grid search used to identify classical autoencoders.}

\vspace{0.3cm}
\begin{ruledtabular}
\begin{tabular}{cc}
Hyperparameter & Values\\
\hline
Batch Sizes & [50,500,1000] \\
Hidden Layers & range[1,10] \\
epochs & [40,60,80] \\
Latent Space & range[1,5] \\
\end{tabular}
\end{ruledtabular}
\caption{Details of the grid search used to identify quantum autoencoders seen in Figure \ref{fig:qcd_roc}.}
\end{table}

\end{document}